\documentclass[preprint,showpacs,preprintnumbers,amsmath,amssymb]{revtex4-2}
\usepackage{bm}
\usepackage{amsmath}
\usepackage{amssymb}
\usepackage{amsfonts}
\usepackage{graphicx}
\begin{document}

\title{On inclusion of the long-range proton-proton Coulomb force
in the three-nucleon scattering Faddeev calculations}

\author{H.~Wita{\l}a}
\affiliation{M. Smoluchowski Institute of Physics, Jagiellonian
University,
                    PL-30059 Krak\'ow, Poland}

\author{J.~Golak}
\affiliation{M. Smoluchowski Institute of Physics, Jagiellonian
University,
PL-30059 Krak\'ow, Poland}

\author{R.~Skibi\'nski}
\affiliation{M. Smoluchowski Institute of Physics, Jagiellonian
University,
                    PL-30059 Krak\'ow, Poland}

\date{\today}

\begin{abstract}
  We propose a simplified approach to incorporate
  the long-range proton-proton (pp)
Coulomb force in the three-nucleon (3N) scattering calculations,
 based on exact formulation presented in
 Eur. Phys. Journal A {\bf{41}}, 369 (2009) and {\bf{41}}, 385 (2009).  
 It permits us to get elastic proton-deuteron (pd) scattering and
 breakup  observables relatively simply 
 by performing standard Faddeev calculations as known for the neutron-deuteron
 (nd) system.
 The basic ingredient in that approach is a 3-dimensional screened pp  Coulomb
 t-matrix obtained by  numerical solution of the 3-dimensional
Lippmann-Schwinger (LS) equation.
 Based on this t-matrix
pure Coulomb transition terms  contributing to 
elastic scattering and breakup are calculated without any need for partial wave
decomposition. For elastic scattering such a term  removes the
Rutherford amplitude for point deuteron proton-deuteron (pd)
scattering. For breakup it has never been applied
in spite of the fact that its contributions could become important in some
 regions of the breakup phase space. 
 We demonstrate numerically that the pd elastic observables
 can be determined directly from the resulting
 3N amplitudes without any renormalization, simply by
 increasing the screening radius in order to reach the
 existing screening limit. However, 
 for pd breakup the renormalization of the contributing
 on-shell amplitudes is required. We apply our approach 
 in a wide energy range of the incoming proton for pd elastic scattering as
 well as for pd breakup reaction.
\end{abstract}


\maketitle \setcounter{page}{1}

\section{Introduction}
 \label{intro}

 For a long time the problem how to include the Coulomb force in 
 the analysis  of nuclear
 reactions with more than two nucleons has attracted wide attention.
 The main difficulty  is the  long-range nature of the Coulomb force which
prevents the application of the standard techniques developed for
short-range interactions. One possible way to 
 include the Coulomb force is to use a screened
Coulomb interaction and to reach the pure Coulomb limit through
application of a renormalization procedure~\cite{Alt78,Alt96,Alt94,Alt2002}.

The high quality of the available pd data for elastic 
 scattering and for the deuteron breakup reaction below the pion
production threshold requires a theoretical
analysis with  the pp Coulomb force included in the calculations
performed with modern nuclear forces. For
this 3N system using the Faddeev scheme high-precision numerical
predictions for different observables in  both processes have been 
obtained \cite{physrep96}, however, only under the restriction to
short-ranged nuclear interactions.

First results for the elastic pd
scattering with modern nuclear forces and the
Coulomb force included, were provided in a variational
hyperspherical harmonic approach~\cite{kievski}. The
 inclusion of the Coulomb force became possible in addition to elastic pd
scattering also for the pd
breakup reaction~\cite{delt2005br}.
 In ~\cite{kievski} the exact Coulomb force in coordinate
representation was used directly.
Contrary to that in~\cite{delt2005br} a screened pp Coulomb force
 was applied in momentum space and in a partial wave basis. In
order to get the final predictions which can be compared to the
data, the limit to the unscreened situation was taken 
numerically, applying a renormalization to the resulting
3N on-shell amplitudes~\cite{delt2005el,delt2005br,delt2006br}.
This allowed the authors for the
 first time to analyze high-precision pd breakup data together with higher
 energy pd elastic scattering ones, and provided
a significant improvement of data description in the cases where the
Coulomb force plays an important role, see e.g. ~\cite{stephan}.

 In spite of the substantial progress in the pd breakup treatment 
achieved in~\cite{delt2005el,delt2005br,delt2006br}
 some important
 questions remained unanswered.
 One is the inability to understand the pp quasi-free-scattering (QFS) and 
 pd space-star (SST) cross sections (see Introduction in~\cite{elascoul}).  
 It motivated us to reconsider the  inclusion of the Coulomb force
in momentum space Faddeev calculations. The main concern in such
type of calculations is the application of a partial wave decomposition
to the long-ranged  Coulomb force. Even when screening is applied, 
it  seems reasonable to treat from the beginning the screened pp
Coulomb t-matrix without partial wave decomposition, because the required
limit
 of vanishing screening leads necessarily to  a drastic increase
of the number of partial wave states involved, what in consequence 
 makes the 
number of 3N partial waves required for convergence extremely large. 
  That fact prompted us to develop in~\cite{elascoul} a novel approach 
to incorporate the pp Coulomb
force in the momentum space 3N Faddeev calculations. 
 It is based on a standard
 formulation for short range forces and relies on the screening  of the
long-range Coulomb interaction. In order to avoid  all uncertainties
connected with the application of the partial wave expansion, 
 inevitable  when working
with long-range forces, we used directly the 3-dimensional pp screened
Coulomb t-matrix.
We demonstrated the 
feasibility of that approach in the case of elastic pd scattering 
using a simple dynamical model for the nuclear part of the interaction. 
 It turned out that the screening limit exists without the need of 
renormalization not only for pd elastic scattering 
observables but for the elastic pd amplitude itself. 
In \cite{elascoul} we demonstrated 
that the physical pd elastic  scattering amplitude can be obtained from
the off-shell solutions of the Faddeev equation and has a well
defined screening limit.

In  \cite{brcoul} we extended that  approach to 
 pd breakup. Again 
 we applied directly the 3-dimensional screened pp Coulomb
t-matrix without relying on a partial wave decomposition. In 
contrast to elastic scattering, where the amplitude itself 
 does not require renormalization, in the case of  pd breakup the
on-shell solutions of the Faddeev equation are required, which  
 necessitates renormalization in the screening limit.

 Another problem concerns the treatment throughout the  calculations
 of the pp Coulomb
interaction in its proper coordinate, which is the relative
proton-proton distance. Any
deviation from this restriction can cause effects which are
difficult to estimate. In our formulation both for elastic pd scattering
and breakup the pp Coulomb interaction is treated in correct coordinates
giving for both processes the Rutherford contribution to the transition
amplitudes in a natural way. In all other 
existing calculations such a term was artificially introduced for
elastic pd scattering by taking the point deuteron pd Coulomb amplitude
 and neglecting it altogether for the breakup reaction.

 The main purpose of present investigation is to establish a relatively
 fast and simple calculational scheme
 for getting reliable estimation
 of the pp Coulomb force effects in pd reactions, 
 which will be with respect to the required time and computer resources 
 comparable to the standard nd Faddeev calculations. In our exact approach
 of Refs.~\cite{elascoul} and \cite{brcoul} 
 the need to calculate  a  number of complex terms, whose determination 
 is very demanding with respect to the required computer time and resources,
 makes this approach  unhandy.  
   In view of coming challenges, such as e.g. fine tuning
  of chiral forces by using high precision pd data, 
  a simpler and faster method for calculation of pd reactions is required.

In section \ref{form} for the convenience of the reader  we briefly 
present  the main points of the formalism outlined in detail  
in \cite{elascoul,brcoul} and introduce our simplified calculational scheme. 
The numerical results for pd elastic scattering are presented and discussed
in section \ref{elastic} and for breakup
in section  \ref{breakup}. 
 The summary and conclusions are given in section \ref{sumary}.

\section{Screened pp Coulomb force in Faddeev equations}
\label{form}

In the following we describe our simplified treatment, which arise from
the exact
formulation of Refs. ~\cite{elascoul,brcoul}. For the convenience of
the reader we  repeat the main steps of ~\cite{elascoul,brcoul}. 

We regard the 3N pd system  in the isospin basis where
the 2-body isospin $t$ together with the isospin $ \frac{1}{2}$ of the
third particle  is coupled to the total isospin  $T$ and the
corresponding completeness relation is (the convention  assumed is that
 the proton (neutron) has the isospin magnetic quantum number
 $ - \frac{1}{2}  (\frac{1}{2})$ ):
\begin{eqnarray}
\sum_{tT} | ( t 1/2 )T -1/2> < ( t 1/2) T -1/2| = 1 ~.
\label{1}
\end{eqnarray}.

We use the Faddeev equation in the form when nucleons interact
with pairwise forces only ~\cite{physrep96}
\begin{eqnarray} T| \Phi> =  t P | \Phi> +  t P G_0 T|
  \Phi> ~,
  \label{2}
\end{eqnarray}
 where $ P $ is defined in terms of transposition operators,
 $ P =P_{12} P_{23} + P_{13} P_{23} $, $G_0$ is the free 3N propagator,
 $ |\Phi>$ the initial state composed of a deuteron state and a momentum
 eigenstate of the proton. 
 The t-matrix $t$ is a solution of the 2-body LS equation, 
\begin{eqnarray}
  t = V + V G_0  t ~,
\label{3}
\end{eqnarray}
with the interaction V containing the neutron-proton (np) ($V_{np}^{t0}= < t 0| V | t0>$) and
pp ($V_{pp}^{1 -1}   =    < 1 -1 | V | 1 -1>$) potentials ~\cite{wit91}.
 The pp interaction decomposes into the strong and the
 pure Coulomb part (assumed to be screened and parametrised
 by some parameter $R$) 
\begin{eqnarray}
V_{pp}^{1 -1} = V_{pp}^{strong} +V_{pp}^{c R}\label{7} ~.
\end{eqnarray}
Since the presence of the pp Coulomb force induces large  charge independence
breaking, which leads necessarily
to  coupling of $ T=1/2 $ and $ T=3/2$ states ~\cite{wit91},
  the  complete treatment 
 of the Coulomb force  requires  states with both total isospin values.

Knowing $T| \Phi>$ the breakup as well as the elastic pd scattering
amplitudes
 can be gained by quadratures in the standard manner~\cite{physrep96}. 
We solve Faddeev equations in our  momentum space partial 
 wave basis $|pq\alpha>$
\begin{eqnarray}
|p q \alpha> \equiv  |pq(ls)j(\lambda
\frac {1} {2})I (jI)J (t \frac {1} {2})T>\label{51} ~,
\end{eqnarray}
distinguishing between the partial wave states $|pq\alpha>$ with
total 2N angular momentum $j$ below some value $j_{max}$: $j \le
j_{max}$,
 in which the nuclear, $V_N$, as well as the pp screened Coulomb
interaction, $V_c^R$
 (in isospin $t=1$ states only), act,  and the
states $|pq\beta>$ with $j > j_{max}$, for which only $V_c^R$ acts 
 in the pp subsystem. The states $|pq\alpha>$ and
$|pq\beta>$ form a complete system of states
\begin{equation}
\int {p^2 dpq^2 dq} (\sum\limits_\alpha  {\left| {pq\alpha }
\right\rangle } \left\langle {pq\alpha } \right| +
\sum\limits_\beta  {\left| {pq\beta } \right\rangle } \left\langle
{pq\beta } \right|) = {\rm I} \\ .~ \label{52}
\end{equation}
Projecting Eq.~(\ref{2}) for $ T| \Phi >$ on the $|pq\alpha>$ and
$|pq\beta>$ states one gets the following system of coupled
integral equations
\begin{eqnarray}
 \left\langle {pq\alpha } \right|T\left| {\Phi } \right\rangle  &=&
 \left\langle {pq\alpha } \right|t_{N + c}^R P\left| {\Phi } \right\rangle
\cr
 &+& \left\langle {pq\alpha } \right|t_{N + c}^R PG_0 \sum\limits_{\alpha '}
 {\int {p'^2 dp'q'^2 dq'\left| {p'q'\alpha '} \right\rangle \left\langle
 {p'q'\alpha '} \right|} } T\left| {\Phi } \right\rangle \cr
 &+& \left\langle {pq\alpha } \right|t_{N + c}^R PG_0 \sum\limits_{\beta '}
 {\int {p'^2 dp'q'^2 dq'\left| {p'q'\beta '} \right\rangle \left\langle
 {p'q'\beta '} \right|} } T\left| {\Phi } \right\rangle  \label{53} ~, \\
 \left\langle {pq\beta } \right|T\left| {\Phi } \right\rangle  &=&
 \left\langle {pq\beta } \right|t_c^R P\left| {\Phi } \right\rangle \cr
  &+& \left\langle {pq\beta } \right|t_c^R PG_0 \sum\limits_{\alpha '}
  {\int {p'^2 dp'q'^2 dq'\left| {p'q'\alpha '} \right\rangle
  \left\langle {p'q'\alpha '} \right|} } T\left| {\Phi } \right\rangle \cr
  &+& \left\langle {pq\beta } \right|t_c^R PG_0 \sum\limits_{\beta '}
  {\int {p'^2 dp'q'^2 dq'\left| {p'q'\beta '} \right\rangle
  \left\langle {p'q'\beta '} \right|} } T\left| {\Phi } \right\rangle ~,
 \label{54}
\end{eqnarray}
where $t_{N+c}^R$ and $t_c^R$ are t-matrices generated by
the interactions $V_N+V_c^R$ and $V_c^R$, respectively.
 For states $|\alpha >$ with two-nucleon subsystem isospin 
$t=1$ the corresponding t-matrix element 
$<p\alpha |t_{N+c}^R(E-\frac {3} {4m}q^2)|p'\alpha'>$ is a linear 
combination of the pp, $t_{pp+c}^R$, and the neutron-proton (np), 
$t_{np}$, $t=1$ t-matrices, which are generated by the interactions 
$V_{pp}^{strong}+V_c^R$ and $V_{np}^{strong}$, 
respectively. The 
coefficients of that combination depend on the total isospin $T$ and
$T'$  
of states $|\alpha >$ and $|\alpha' >$~\cite{elascoul,wit91}:
\begin{eqnarray}
<t=1T=\frac{1} {2} |t_{N+c}^R|t'=1T'=\frac{1}
{2}> 
& =&  \frac {1} {3} t_{np} + \frac {2} {3} t_{pp+c}^{R} \cr
<t=1T=\frac{3} {2} |t_{N+c}^R|t'=1T'=\frac{3}
{2}> 
& =&  \frac {2} {3} t_{np} + \frac {1} {3} t_{pp+c}^{R} \cr
<t=1T=\frac{1} {2} |t_{N+c}^R|t'=1T'=\frac{3}
{2}> 
& =&  \frac {\sqrt{2}} {3}( t_{np} - t_{pp+c}^{R}) \cr
<t=1T=\frac{3} {2} |t_{N+c}^R|t'=1T'=\frac{1}
{2}> 
& =&  \frac {\sqrt{2}} {3}( t_{np} - t_{pp+c}^{R})    ~.
\label{8}
\end{eqnarray}
For isospin $t=0$, where  
$T=T'=\frac {1} {2}$:
\begin{eqnarray}
<t=0T=\frac{1} {2} |t_{N+c}^R|t'=0T'=\frac{1}
{2}> 
& =& t_{np} ~.
\label{8a}
\end{eqnarray}
In the case of $t_c^R$ only the screened pp Coulomb force $V_c^R$  acts.

The third term  on the right hand side of (\ref{54}) is
proportional to $$<pq\beta|t_c^RPG_0|p'q'\beta'><p'q'\beta'|t_c^ R ~.$$ A direct
calculation shows that it vanishes, independently of the
value of the total isospin $T$.

Inserting $<pq\beta|T|\Phi>$ from (\ref{54}) into (\ref{53}) one
gets
\begin{eqnarray}
 \left\langle {pq\alpha } \right|T\left| {\Phi } \right\rangle  &=&
 \left\langle {pq\alpha } \right|t_{N + c}^R P\left| {\Phi } \right\rangle
 + \left\langle {pq\alpha } \right|t_{N + c}^R PG_0 t_c^R P\left| {\Phi }
\right\rangle \cr
  &-& \left\langle {pq\alpha } \right|t_{N + c}^R PG_0 \sum\limits_{\alpha
'}
  {\int {p'^2 dp'q'^2 dq'\left| {p'q'\alpha '} \right\rangle
  \left\langle {p'q'\alpha '} \right|} } t_c^R P\left| {\Phi } \right\rangle
\cr
  &+& \left\langle {pq\alpha } \right|t_{N + c}^R PG_0 \sum\limits_{\alpha
'}
  {\int {p'^2 dp'q'^2 dq'\left| {p'q'\alpha '} \right\rangle
  \left\langle {p'q'\alpha '} \right|} } T\left| {\Phi }
  \right\rangle \cr
  &+& \left\langle {pq\alpha } \right|t_{N + c}^R PG_0 t_c^R PG_0
\sum\limits_{\alpha '}
  {\int {p'^2 dp'q'^2 dq'\left| {p'q'\alpha '} \right\rangle
  \left\langle {p'q'\alpha '} \right|} } T\left| {\Phi } \right\rangle \cr
  &-& \left\langle {pq\alpha } \right|t_{N + c}^R PG_0  \sum\limits_{\alpha
'}
  {\int {p'^2 dp'q'^2 dq'\left| {p'q'\alpha '} \right\rangle
  \left\langle {p'q'\alpha '} \right|} } t_c^R PG_0 \cr
  && \sum\limits_{\alpha'' }
  {\int {p''^2 dp''q''^2 dq''\left| {p''q''\alpha'' } \right\rangle
  \left\langle {p''q''\alpha'' } \right|} } T\left| {\Phi } \right\rangle ~.
  \label{55}
 \end{eqnarray}
This is a coupled set of integral equations in the space of only the states
$\left| {{\alpha } } \right\rangle$, which exactly
incorporates the contributions of
the pp Coulomb interaction from all partial wave states up to
infinity.
It can be solved by iteration and Pad\'e
summation. However, the very time consuming and complicated calculation 
 of contributing terms containing the 3-dimensional screened Coulomb t-matrix  
 (the second- and fith-term) prevents solution of that equation
 for  practically interesting case of sufficiently
 large partial wave basis.
 It is our aim in the present investigation to simplify 
  (\ref{55}) without losing its physical content, so that the
 resulting equation will be manageable  as for the nd system.
 
 Actually a glimpse at Eq.~(\ref{55}) reveals a possibility to avoid completely 
 a  calculation of these complicated terms with 3-dimensional Coulomb t-matrix
 and to omit the second-, third- as well as the fifth- and sixth-term
 altogether. Namely, at a specific value of the screening radius R, 
 a finite set of partial waves provides an exact
 reproduction of the 3-dimensional Coulomb t-matrix $t_c^R$. Extending
 the set $\left| {{\alpha } } \right\rangle$ to such a set of states 
 by adding a finite number of channels with higher angular
 momenta, in which only the pp Coulomb interaction is present, permits to
 completely neglect the above mentioned four terms due to their mutual
 cancellation: second with the third and fifth with the sixth term.
 The set (\ref{55}) is then reduced to the identical
 form as in  the nd case
\begin{eqnarray}
 \left\langle {pq\alpha } \right|T\left| {\Phi } \right\rangle  &=&
 \left\langle {pq\alpha } \right|t_{N + c}^R P\left| {\Phi } \right\rangle
\cr
 &+& \left\langle {pq\alpha } \right|t_{N + c}^R PG_0 \sum\limits_{\alpha '}
 {\int {p'^2 dp'q'^2 dq'\left| {p'q'\alpha '} \right\rangle \left\langle
 {p'q'\alpha '} \right|} } T\left| {\Phi } \right\rangle  ~ .
 \label{63} 
\end{eqnarray} 
With an increasing R-value
 the above cancellation  requires more and more partial waves so generally 
 for  some chosen finite set $\left| {{\alpha } } \right\rangle$
 only a partial cancellation is expected. 

 Even in the case  when channels $\left| {{\alpha } } \right\rangle$
 are those in
 which nuclear and pp Coulomb forces  act and only partial cancellation
 occurs, an additional argument prompt
 one to simplify the set (\ref{53})-(\ref{54}). 
 Namely, in Eqs.~(\ref{53}) and (\ref{54}) the strength of the 
coupling between amplitudes
$\left\langle {pq\alpha } \right|T\left| {\Phi } \right\rangle$ and
$\left\langle {pq\beta } \right|T\left| {\Phi } \right\rangle$ is determined
by matrix elements of the permutation operator
$\left\langle {pq\alpha } \right| P \left| {{p'q'\beta } } \right\rangle$.
 Since channels $\left| {{\beta } } \right\rangle$
have the values of the total two-body subsystem angular momentum $j$
larger than the channels $\left| {{\alpha } } \right\rangle$ the
matrix element of the permutation operator between these states
 is smaller than between $\left| {{\alpha } } \right\rangle$ states.
 One can thus argue that the 
 third term in  equation (\ref{53}) and second in (\ref{54}) 
 are  small compared to  the leading terms.
 Neglecting them  would lead again to the set (\ref{63}).

 For a restricted basis $\left| {{\alpha } } \right\rangle$
 ($j \le 3$) it is possible to compute
 the second term in (\ref{55}) within a reasonable amount of computer time and
 resources. Therefore      
 we would  also like to look at these cancellations in a more direct way.   
 By omitting only the second term in Eq.~(\ref{54}) one 
 reduces (\ref{55}) to a form containing the first pair of leading terms
 with the 3-dimensional and its partial-wave decomposed counterpart
 Coulomb t-matrix
\begin{eqnarray}
 \left\langle {pq\alpha } \right|T\left| {\Phi } \right\rangle  &=&
 \left\langle {pq\alpha } \right|t_{N + c}^R P\left| {\Phi } \right\rangle
 + \left\langle {pq\alpha } \right|t_{N + c}^R PG_0 t_c^R P\left| {\Phi }
\right\rangle \cr
  &-& \left\langle {pq\alpha } \right|t_{N + c}^R PG_0 \sum\limits_{\alpha
'}
  {\int {p'^2 dp'q'^2 dq'\left| {p'q'\alpha '} \right\rangle
  \left\langle {p'q'\alpha '} \right|} } t_c^R P\left| {\Phi } \right\rangle
\cr
  &+& \left\langle {pq\alpha } \right|t_{N + c}^R PG_0 \sum\limits_{\alpha
'}
  {\int {p'^2 dp'q'^2 dq'\left| {p'q'\alpha '} \right\rangle
  \left\langle {p'q'\alpha '} \right|} } T\left| {\Phi }
  \right\rangle  ~.
  \label{65}
 \end{eqnarray}

We will solve both equations (\ref{63}) and (\ref{65}) and demonstrate, 
how cancellation effects between  the second and fifth term in (\ref{55})
affect  the elastic scattering and breakup observables.

 After solving 3N Faddeev equations   the transition amplitude
  for elastic scattering is  given
by~\cite{gloeckle83,physrep96}
\begin{eqnarray}
 \left\langle {\Phi '} \right|U\left| {\Phi } \right\rangle
 &=&
 \left\langle {\Phi '} \right|PG_0^{ - 1}  + PT\left| {\Phi }
 \right\rangle ~.
\label{76}
 \end{eqnarray}
The first contribution  is independent of  the pp Coulomb
force and can be calculated without partial wave decomposition using expression
given  in Appendix C of Ref.~\cite{elascoul}. 
 To calculate the second contribution in (\ref{76}) one needs
$\left\langle {\vec p\vec q~} \right|T\left| {\Phi }
\right\rangle$ composed of low ($\alpha$) and high ($\beta$)  partial wave 
contributions for $ T | \Phi>$. Using the completeness relation
(\ref{52}) one gets:
\begin{eqnarray}
&&\left\langle {\vec p\vec q~} \right|T\left| {\Phi }
\right\rangle  = \left\langle {\vec p\vec q~}
\right|\sum\limits_{\alpha '} {\int {p'^2 dp'q'^2 dq'\left|
{p'q'\alpha '} \right\rangle \left\langle {p'q'\alpha '} \right|}
} T\left| {\Phi } \right\rangle \cr && - \left\langle {\vec
p\vec q~} \right|\sum\limits_{\alpha '} {\int {p'^2 dp'q'^2
dq'\left| {p'q'\alpha '} \right\rangle \left\langle {p'q'\alpha '}
\right|} } t_c^R P\left| {\Phi } \right\rangle \cr &&-
\left\langle {\vec p\vec q~} \right|\sum\limits_{\alpha '} {\int
{p'^2 dp'q'^2 dq'\left| {p'q'\alpha '} \right\rangle \left\langle
{p'q'\alpha '} \right|} } t_c^R PG_0
 \sum\limits_{\alpha
''} {\int {p''^2 dp''q''^2 dq''\left| {p''q''\alpha ''}
\right\rangle \left\langle {p''q''\alpha ''} \right|} } T\left|
{\Phi } \right\rangle \cr && + \left\langle {\vec p\vec q~}
\right|t_c^R P\left| {\Phi } \right\rangle  + \left\langle
{\vec p\vec q~} \right|t_c^R PG_0 \sum\limits_{\alpha '} {\int
{p'^2 dp'q'^2 dq'\left| {p'q'\alpha '} \right\rangle \left\langle
{p'q'\alpha '} \right|} } T\left| {\Phi } \right\rangle ~.
\label{78}
\end{eqnarray}
It follows, that in addition to the amplitudes $<pq\alpha|T|\Phi >$ also the
partial wave projected amplitudes $<pq\alpha|t_c^RP|\Phi >$ and
 $<pq\alpha|t_c^RPG_0|\alpha'><\alpha'|T|\Phi>$ are required. 
The expressions for the contributions of these three terms to the
transition amplitude for elastic scattering (and breakup)   are
given in Appendix B of Ref. ~\cite{elascoul}. 
The last  two terms in (\ref{78}) must    be calculated using
directly the $3$-dimensional screened Coulomb t-matrices.  Expressions
for  $\left\langle {\vec p\vec q~}\right|t_c^R P\left| {\Phi } \right\rangle$
 (breakup) and
$\left\langle {\Phi~'} \right|Pt_c^R P\left| {\Phi } \right\rangle$
 (elastic scattering)  are  given in Appendix C of Ref. ~\cite{elascoul}. 
The term $\left\langle {\Phi~'} \right|Pt_c^R P\left| {\Phi }\right\rangle$
 for large values of the 
 screening radius $R$ is the contribution from the pp  Coulomb force to the pd
 scattering, which quite often  was replaced by
 the Rutherford amplitude for point deuteron Coulomb pd scattering.
  To the best of our knowledge the analogous term for breakup,
  $\left\langle {\vec p\vec q~}\right|t_c^R P\left| {\Phi } \right\rangle$,  
  was altogether omitted in all pd breakup calculations. 
 The expression for the last matrix element $<\vec p \vec q~
 |t_c^RPG_0|\alpha '><\alpha '|T|\Phi>$ is given in Appendix  D
 of Ref. ~\cite{elascoul}. It provides a correction to the pure Coulomb term in
 pd elastic scattering and breakup due to the strong interactions
  between nucleons.
 It is interesting to note that all  terms containing 3-dimensional
 Coulomb t-matrix, both for elastic scattering and breakup, can be calculated
 using analytical expressions for the screening limit of that matrix.

 Since we  restrict ourselves 
 to approximations (\ref{63})
 and (\ref{65}), it would seem validated to 
 omit  the third and fifth term  
 in   Eq.~(\ref{78}) reducing it to 
\begin{eqnarray}
&&\left\langle {\vec p\vec q~} \right|T\left| {\Phi }
\right\rangle  = \left\langle {\vec p\vec q~}
\right|\sum\limits_{\alpha '} {\int {p'^2 dp'q'^2 dq'\left|
{p'q'\alpha '} \right\rangle \left\langle {p'q'\alpha '} \right|}
} T\left| {\Phi } \right\rangle \cr && - \left\langle {\vec
p\vec q~} \right|\sum\limits_{\alpha '} {\int {p'^2 dp'q'^2
dq'\left| {p'q'\alpha '} \right\rangle \left\langle {p'q'\alpha '}
\right|} } t_c^R P\left| {\Phi } \right\rangle \cr
            && + \left\langle {\vec p\vec q~}
\right|t_c^R P\left| {\Phi } \right\rangle ~.
\label{79}
\end{eqnarray}

Actually, there is no justification for rejecting these two pure Coulomb terms.
 Contrary, it seems
unavoidable that the pure Coulomb terms must receive contributions from
strong interactions between nucleons as indicated by the second term in
(\ref{54}), importance of which will probably
 depend on the energy. However, in the present investigation 
 we stick  first to approximation (\ref{79}) for the transition amplitude,
 deferring the
 problem of importance of rejected terms for a  later stage of the present
 study.

The transition amplitude for 
 breakup $<\Phi_0|U_0|\Phi>$ is given
in terms of $T\left| {\Phi } \right\rangle$ by~\cite{gloeckle83,physrep96}
\begin{eqnarray}
 \left\langle {\Phi _0 } \right|U_0 \left| {\Phi } \right\rangle  &=&
 \left\langle {\Phi _0 } \right|(1 + P)T\left| {\Phi } \right\rangle ~,
\label{56}
 \end{eqnarray}
where $ | \Phi_0> = | \vec p
\vec q m_1 m_2 m_3 \nu_1 \nu_2 \nu_3 > $ is the  state of three free
outgoing nucleons. 
 The permutations acting in momentum-, spin-, and isospin-spaces 
can be applied to the bra-state $< \phi_0| = < \vec p
\vec q m_1 m_2 m_3 \nu_1 \nu_2 \nu_3 | $, changing the sequence of 
nucleons spin and isospin magnetic
quantum numbers $ m_i$ and  $\nu_i$ and leading to well known linear
combinations of the Jacobi momenta $ \vec p,\vec q$. Thus
evaluating (\ref{56}) it is sufficient to regard the general
amplitudes $< \vec p \vec q m_1 m_2 m_3 \nu_1 \nu_2 \nu_3 | 
T\left| {\Phi }\right\rangle \equiv \left\langle {\vec p\vec q~} 
\right|T\left| {\Phi }
\right\rangle $, which are given again by (\ref{79}). Also 
 for breakup, in addition to the amplitudes $<pq\alpha|T|\Phi >$,  the
partial wave projected amplitude $<pq\alpha|t_c^RP|\Phi >$ is required. The
expressions for the contributions of these two terms to the
transition amplitude for the breakup reaction are
given in Appendix B of Ref.~\cite{elascoul}.
 The last term in (\ref{79}) must be calculated using
directly the $3$-dimensional screened Coulomb t-matrix.
It corresponds to the Rutherford amplitude in elastic pd scattering. 
In Appendix C  
of Ref.~\cite{elascoul} the expression for 
$\left\langle {\vec p\vec q~} \right|t_c^R
P\left| {\Phi } \right\rangle$  is provided.

The  Faddeev equations (\ref{63}) and
 (\ref{65}) are 
  well defined for any finite screening radius.  The
important challenge is to control the screening  limit for the
physical pd elastic scattering and breakup  amplitudes (\ref{79}).
In the case of elastic
scattering we provided in Ref.~\cite{elascoul} arguments and showed 
 numerically that the physical elastic pd scattering amplitude itself 
 has a well 
defined screening limit and does not require renormalization. 
This can be traced back to the fact that in order
 to get the elastic pd scattering 
amplitude it is sufficient to solve the Faddeev 
equations (\ref{63})  ( (\ref{65}) ) for off-shell values of the Jacobi momenta 
\begin{eqnarray}
    \frac {p^2} {m} + \frac{3}{4m} q^2 \ne  E ~.
\label{73}
\end{eqnarray}
The off-shell Faddeev amplitudes 
 $\left\langle {pq\alpha } \right|T\left| {\Phi } \right\rangle$ of 
Eqs.~(\ref{63})  ( (\ref{65}) )
 are determined by off-shell nucleon-nucleon t-matrix elements 
$t(p,p';E-\frac {3} {4m}q^2)$, which have a well defined screening limit 
(see ~\cite{elascoul} as well as discussion and examples in ~\cite{brcoul}). 
 Also the  off-shell 3-dimensional pure Coulomb t-matrix is known analytically
 (see \cite{chen72,kok1980} and references therein)
\begin{eqnarray}
  \left\langle {\vec p~'} \right| t_c^R(\frac {k^2} {m})
  \left| {\vec p} \right\rangle \to \frac {e^2} {2\pi^2}
  \frac {1+I(x)} {({\vec p~'}-{\vec p})^2}
  \label{73a}
\end{eqnarray}
with $$I(x)=\frac {1} {x} [_2F_1(1,i\eta;1+i\eta;\frac {x+1} {x-1})
  - _2F_1(1,i\eta;1+i\eta;\frac {x-1} {x+1})]$$ and
$$x^2=1+\frac {(p'^2-k^2)(p^2-k^2)} {k^2({\vec p~'}-{\vec p})^2}~,$$
 where $_2F_1$ is a hypergeometric function \cite{abr_steg}. 
Thus also the  Coulomb term
 $\left\langle {\Phi}' \right| Pt_c^RP 
\left| {\Phi} \right\rangle$ has a well defined
screening limit \cite{skib2009}.  

Contrary to  pd elastic scattering the physical breakup 
amplitude (\ref{56}) corresponds to the on-shell values of Jacobi momenta 
\begin{eqnarray}
    \frac {p^2} {m} + \frac{3}{4m} q^2 =  E \equiv \frac{3}{4m} q_{max}^2 ~.
\label{74}
\end{eqnarray}
That means that the physical pd breakup amplitude (\ref{56}) requires 
on-shell Faddeev amplitudes 
$\left\langle {p_0q\alpha } \right|T\left| {\Phi } \right\rangle$ together with
 the two, also on-shell, additional terms in (\ref{79}), with 
 $p_0=\sqrt {\frac {3} {4} (q_{max}^2-q^2)}$.
 The on-shell Faddeev amplitudes can be obtained from the off-shell solutions 
$\left\langle {pq\alpha } \right|T\left| {\Phi } \right\rangle$ using 
 (\ref{55}):
\begin{eqnarray}
 \left\langle {p_0q\alpha } \right|T\left| {\Phi } \right\rangle  &=&
 \left\langle {p_0q\alpha } \right|t_{N + c}^R P\left| {\Phi } \right\rangle
 + \left\langle {p_0q\alpha } \right|t_{N + c}^R PG_0 t_c^R P\left| {\Phi }
\right\rangle \cr
  &-& \left\langle {p_0q\alpha } \right|t_{N + c}^R PG_0 \sum\limits_{\alpha
'}
  {\int {p'^2 dp'q'^2 dq'\left| {p'q'\alpha '} \right\rangle
  \left\langle {p'q'\alpha '} \right|} } t_c^R P\left| {\Phi } \right\rangle
\cr
  &+& \left\langle {p_0q\alpha } \right|t_{N + c}^R PG_0 \sum\limits_{\alpha
'}
  {\int {p'^2 dp'q'^2 dq'\left| {p'q'\alpha '} \right\rangle
  \left\langle {p'q'\alpha '} \right|} } T\left| {\Phi }
  \right\rangle \cr
  &+& \left\langle {p_0q\alpha } \right|t_{N + c}^R PG_0 t_c^R PG_0
\sum\limits_{\alpha '}
  {\int {p'^2 dp'q'^2 dq'\left| {p'q'\alpha '} \right\rangle
  \left\langle {p'q'\alpha '} \right|} } T\left| {\Phi } \right\rangle \cr
  &-& \left\langle {p_0q\alpha } \right|t_{N + c}^R PG_0  \sum\limits_{\alpha
'}
  {\int {p'^2 dp'q'^2 dq'\left| {p'q'\alpha '} \right\rangle
  \left\langle {p'q'\alpha '} \right|} } t_c^R PG_0 \cr
  && \sum\limits_{\alpha'' }
  {\int {p''^2 dp''q''^2 dq''\left| {p''q''\alpha'' } \right\rangle
  \left\langle {p''q''\alpha'' } \right|} } T\left| {\Phi } \right\rangle ~. 
  \label{eq55a}
 \end{eqnarray}

This reduces for the case of  (\ref{63}) to:
\begin{eqnarray}
 \left\langle {p_0q\alpha } \right|T\left| {\Phi } \right\rangle  &=&
 \left\langle {p_0q\alpha } \right|t_{N + c}^R P\left| {\Phi } \right\rangle
\cr
  &+& \left\langle {p_0q\alpha } \right|t_{N + c}^R PG_0 \sum\limits_{\alpha
'}
  {\int {p'^2 dp'q'^2 dq'\left| {p'q'\alpha '} \right\rangle
  \left\langle {p'q'\alpha '} \right|} } T\left| {\Phi }
  \right\rangle    ~, 
  \label{eq80}
\end{eqnarray}
or for  (\ref{65}) to:
\begin{eqnarray}
 \left\langle {p_0q\alpha } \right|T\left| {\Phi } \right\rangle  &=&
 \left\langle {p_0q\alpha } \right|t_{N + c}^R P\left| {\Phi } \right\rangle
 + \left\langle {p_0q\alpha } \right|t_{N + c}^R PG_0 t_c^R P\left| {\Phi }
\right\rangle \cr
  &-& \left\langle {p_0q\alpha } \right|t_{N + c}^R PG_0 \sum\limits_{\alpha
'}
  {\int {p'^2 dp'q'^2 dq'\left| {p'q'\alpha '} \right\rangle
  \left\langle {p'q'\alpha '} \right|} } t_c^R P\left| {\Phi } \right\rangle
\cr
  &+& \left\langle {p_0q\alpha } \right|t_{N + c}^R PG_0 \sum\limits_{\alpha
'}
  {\int {p'^2 dp'q'^2 dq'\left| {p'q'\alpha '} \right\rangle
  \left\langle {p'q'\alpha '} \right|} } T\left| {\Phi }
  \right\rangle    ~. 
  \label{eq81}
\end{eqnarray}

These on-shell amplitudes together with additional, also on-shell, 
terms in (\ref{79}) define the physical breakup amplitude (\ref{56}). 
That in consequence requires half-shell t-matrix elements 
$t(p_0,p';\frac {p_0^2} {m})$ which are of 3 types:  
the partial wave projected pure screened Coulomb $t_c^R$ generated by 
$V_{c}^R$, the 
 partial wave projected  $t_{N+c}^R$ generated by
 $V_{strong}+V_{c}^R$, 
and the 3-dimensional screened Coulomb t-matrix elements.

It is well known \cite{Alt78,ford1964,ford1966}  that in the screening limit 
 $ R \rightarrow \infty$ such half-shell t-matrices  
acquire an infinitely oscillating phase factor $ e^ {i \Phi_R(p)}$,
 where $\Phi_R(p)$ depends on the type of the screening. 
 For the exponential screening: 
\begin{equation}
V_c^R(r) = \frac{\alpha} {r} e^{-{(\frac {r} {R})}^n} 
\label{eq.2}
\end{equation}
its form  depends
on two parameters, the screening radius $R$ and the power $n$. 
At a given value $n$ the pure Coulomb potential results for $R \rightarrow
\infty$. As has been shown in~\cite{kamada05} based
on~\cite{taylor1,taylor2},  the related phase $ \Phi_n^R(p)$ is given as
\begin{eqnarray}
  \Phi_n^R(p) = -\eta [ ln(2pR) - \frac {\gamma} {n} ] +\eta\sum_{k=1}^{\infty}
  \frac {(-1)^k} {knk!(2pR)^{kn}} ~,
\label{eq.8}
\end{eqnarray}
where  $\gamma=0.5772\dots$ is the 
Euler number and $\eta = \frac{ m_p \alpha}{2
p} $ is the so-called Sommerfeld parameter.

That oscillatory phase factor appearing in the half-shell
proton-proton 
 t-matrices requires 
 a careful treatment  to get 
the screening limit for the  
$\left\langle {p_0q\alpha } \right|T\left| {\Phi } \right\rangle$ 
amplitudes. Namely for the 
states $|\alpha >$ with the two-nucleon subsystem isospin 
$t=1$ the corresponding t-matrix element 
$<p_0\alpha |t_{N+c}^R(\frac {p_0^2} {m})|p'\alpha'>$ is a linear 
combination of the pp and neutron-proton (np) $t=1$ t-matrices,  
 with  coefficients  which  depend on the total isospin T and T' 
of the states $|\alpha >$ and $|\alpha' >$ (see discussion after (\ref{54})).
 It follows that to achieve the screening limit one 
 needs to renormalize breakup amplitudes by removing from them the oscillatory
 phase factor induced by
 the half-shell pp t-matrix $t_{pp+c}^R$. 
 The term in that linear combination coming with the np t-matrix
 $t_{np}$ is not influenced by pp Coulomb force. 
 In \cite{brcoul} we erroneously suggested a renormalization
 in that combination  before performing the action 
 of the operators in (\ref{eq80}) and (\ref{eq81}). This is however incorrect
 since amplitudes obtained in this way do not fulfill the Faddeev equations and
 would lead to false results for breakup observables.
 The only proper place to perform renormalization
 is during calculation of the breakup transition amplitude. 

 The breakup transition amplitude is built 
  up from three contributions due to the action of the $(1+P)$ operator
  in Eq.~(\ref{56}). For a given specification of outgoing nucleons imposed
    by experimental conditions, only one of these contributions corresponds
    to the case that the neutron is a spectator nucleon $1$ and
    two protons form the interacting $2-3$ pair  (pp-partition). 
    When calculating breakup transition amplitude
    the summation over $t=1$ $ | \alpha >$ 
    states with the total 3N isospin $T=\frac {1} {2}$ or  
    $T=\frac {3} {2}$ provides, that in that particular partition
    only the component of the breakup transition
    amplitude driven by pp half-shell
    t-matrix  will be left. Contrary, the two other partitions with the
    proton as a spectator
    nucleon $1$, will contain only component of the breakup transition
    amplitude driven by
    np half-shell t-matrix (np-partitions).

    To be more specific let us consider the contribution to the breakup
    transition amplitude
    from  particular  isospin $t=1$ channels $\alpha$, which differ only
    in their total isospin value $T=\frac {1} {2}$ or $\frac {3} {2}$,
    in a partition defined by
    a spectator nucleon $1$ with isospin projection $\nu_1$ and nucleons
    $2$ and $3$ with isospin projections $\nu_2$ and $\nu_3$, respectively.
    In the following we keep only isospin quantum numbers.
    This contribution is
    proportional to
\begin{eqnarray}
  \sum_{T,T'} C(T) ~ \left\langle \alpha_T \left| t_{N+c}^R \right|
  \alpha_{T'} \right\rangle  \left\langle \alpha_{T'} \right| A \left|
            {\Phi } \right\rangle   ~,
  \label{200}
\end{eqnarray}    
with $A \left| {\Phi } \right\rangle =( P+PG_0T) \left| {\Phi }
\right\rangle $ in case of Eq.~(\ref{eq80}), and 
$$C(T) \equiv \left< \frac {1} {2} \nu_2 \frac {1} {2} \nu_3 |
t=1 \nu_2 + \nu_3 \right>
\left< t=1 \nu_2 + \nu_3 \frac {1} {2} \nu_1 | T \nu_2
+ \nu_3 + \nu_1 \right> ~.$$ 

That gives the contribution to the pp-partition
($\nu_2=\nu_3=-\frac {1} {2}$,
$\nu_1=+\frac {1} {2}$) to be proportional to $ t_{pp+c}^R$:
$$ t_{pp+c}^R \left[ - \sqrt{\frac {2} {3}}
\left\langle T'=\frac {1} {2} \left| A
   \right| {\Phi } \right\rangle  +
   \frac {1} {\sqrt{3}}  \left\langle T'=\frac {3} {2}  \left| A
   \right| {\Phi } \right\rangle  ~ \right] $$
and the contribution to the np-partition ($\nu_2=-\frac {1} {2},
\nu_3=+\frac {1} {2}$ or $\nu_2=+\frac {1} {2}, \nu_3=-\frac {1} {2}, 
 \nu_1=-\frac {1} {2} $) to be proportional to $ t_{np}$:
 $$ t_{np} \left[~ \frac {\sqrt{2}} {2\sqrt{3}}  \left\langle T'=\frac {1} {2}
   \left| A \right| {\Phi } \right\rangle  + 
   \frac {1} {\sqrt{3}}  \left\langle T'=\frac {3} {2}
   \left| A \right| {\Phi } \right\rangle ~ \right] ~. $$  

    That means that renormalization has to  be done  on the level
    of contributing breakup amplitudes  by renormalizing the amplitude  
    $\left\langle {p_0q\alpha } \right|T\left| {\Phi } \right\rangle$ 
  for the pp-partition   with the neutron as  a spectator
  nucleon $1$.
  The same concerns the renormalization of contributions from
 $\left\langle {p_0q\alpha } \right| t_{c}^RP\left| {\Phi } \right\rangle$
 and from the 3-dimensional pp t-matrix
 $\left\langle {\vec p_0 \vec q} \right| t_{c}^RP\left| {\Phi } \right\rangle$,
 whose terms, however,  contribute only
 in a pp-partition with the neutron as  a spectator
 nucleon $1$.  Since the half-shell pure Coulomb t-matrix is analytically given
 by \cite{kok1981}
\begin{eqnarray}
  \left\langle {\vec p~'} \right| t_c^R(\frac {k^2} {m})
  \left| {\vec k} \right\rangle \to C_0 e^{i\sigma_0} \frac {k\eta} {\pi^2q^2}
  ( \frac {p'^2 - k^2} {q^2} )^{i\eta}
  \label{73b}
\end{eqnarray}
with $\vec q=\vec p~' - \vec k$, the pure Coulomb phase
shift $\sigma_0=\Gamma(1+i\eta)$, and Coulomb penetrability 
$C_0^2=\frac {2\pi \eta} {e^{2\pi \eta}-1}$,  
 also the renormalized  term
 $\left\langle \vec p_0 \vec q \right| t_c^{R}P 
 \left| {\Phi} \right\rangle$ has a well defined
 screening limit \cite{skib2009}.  

 Another possibility to get on-shell breakup amplitudes is offered by
 interpolations 
 of the off-shell amplitudes to on-shell values of Jacobi momenta. 
Contrary to the half-shell, the off-shell t-matrix elements do not 
acquire such an 
oscillating phase and their screening limit is well
defined.  
Thus also off-shell  $\left\langle {p'q'\alpha '} \right| T\left| {\Phi }
    \right\rangle$ as well as off-shell 
    $\left\langle {p'q'\alpha '} \right| t^R_cP \left| {\Phi }  \right\rangle$ 
   and a 3-dimensional Coulomb t-matrix  do not 
acquire such an oscillating phase and their screening limits are  well
defined.
 However,  their  half-shell counterparts 
 obtained by interpolation from off-shell to on-shell values of Jacobi momenta 
 will gain  the oscillating phase in their ``pp''-components. 
  These interpolated on-shell amplitudes   must be thus also renormalized.

 The above two ways to get on-shell breakup amplitudes  
 should provide the same results for breakup observables. It  offers an
 additional verification of numerics.

\section{Numerical results}
\label{results}
\subsection{Elastic scattering}
\label{elastic}

We applied the above  approaches 
using a  dynamical model in which three nucleons  interact  with the
AV18  nucleon-nucleon potential ~\cite{av18} restricted
to act only in partial waves with $j \le 3$.
The only reason why we restricted ourselves
 to this rather small  value of $j_{max}=3$,
is that we would like to compare results of our simplified approach (\ref{63}) 
with that of Eq.~(\ref{65}).
 This requires computation of
the second component in the leading term with a 3-dimensional Coulomb t-matrix.
With $j_{max}=3$ the needed computer time and resources
are definitely affordable.   
 The pp Coulomb force was screened exponentially
\begin{eqnarray}
V_{pp}^{cR}(r) &=& \frac{e^2}{r} e^{-({\frac{r} {R}})^n }
\label{scr_1}
\end{eqnarray}
with the screening radius $R$ and $n=4$.

To investigate the screening limit $R \to \infty$ we generated a
set of partial-wave decomposed t-matrices, $t_c^R$, based on the
screened pp Coulomb force alone,  or    combined with the
pp nuclear
interaction, $ t_{ N+C}^ R $, taking $R=5, 10, 20, 30,$ and
$40$~fm. With that dynamical input we solved the  Faddeev
equation (\ref{63}) for the total angular momenta of the p-p-n
system up to $J \le \frac {31} {2}$ and both parities. 
 With the same
screening radii  we generated  the 3-dimensional pure Coulomb
t-matrices $t^R_c$ by solving  the 3-dimensional LS equation.
For $R=40$~fm we solved also the Eq.~(\ref{65}) calculating additional
two leading terms containing 3-dimensional Coulomb t-matrix
as well as its partial wave projected counterpart.

\begin{figure}
  \includegraphics[scale=0.6]{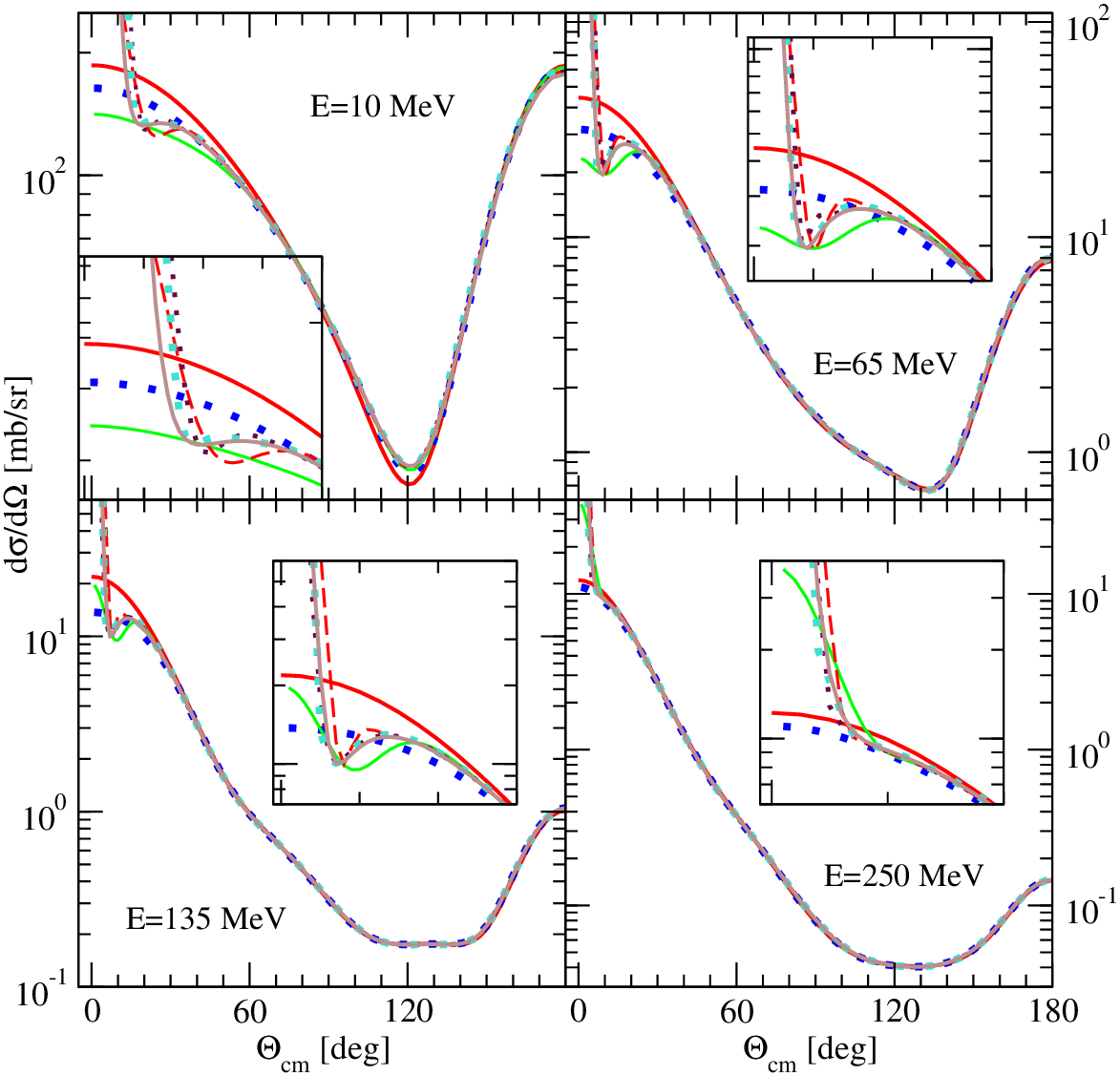}
\caption{(color online) The convergence in the cut-off radius R
of the  pd elastic scattering cross section
$\frac {d\sigma} {d\Omega}$ shown as a function of the c.m. angle
$\Theta_{cm}$ at the incoming lab. proton energy
$E=10, 65, 135,$ and $250$~MeV. These cross sections were
calculated using approach based on Eq.~(\ref{63}), taking elastic scattering
transition amplitude (\ref{79}) 
with the screened Coulomb
force and the AV18 nucleon-nucleon potential~\cite{av18}
restricted to the $j \le 3$ partial waves.
 The screening
 radii are  ($n=4$):  $R=5$~fm (blue dotted line),
 $R=10$~fm (green solid line),
 $R=20$~fm (red long-dashed line),
 $R=30$~fm (maroon dotted line), 
 $R=40$~fm (turquoise dotted line).
 The brown solid line  corresponds to the $R=40$~fm result with
 the pure Coulomb term calculated according to Eq.~(\ref{73a}). 
The red solid line is the 
nd elastic scattering cross section and 
 the insets show the region of small angles.
 }
 \label{fig1}
\end{figure}

In Fig.~\ref{fig1} we show the convergence in the screening radius $R$ of
the pd elastic scattering cross section. The pd predictions of
the first approach (\ref{63}), using elastic scattering transition
amplitude  (\ref{79}) and the Coulomb term
$\left\langle {\Phi}' \right| Pt_c^{R}P\left| {\Phi} \right\rangle$,
 calculated with the corresponding screening
radius $R$, are compared with the nd
 angular distributions at the incoming proton or neutron lab. energies 
$E=10, 65, 135,$ and $250$~MeV. 
 On the scale of the figure the pd cross sections for all screening radii $R$
 are practically
 indistinguishable with exception of forward c.m. angles below $\approx 30^o$, 
 shown in insets. 
 At very forward angles cross sections for $R=5$ and $10$~fm clearly deviate 
 but starting from $R=40$~fm  the screening limit is achieved, with
 exception of $E=10$~MeV, which requires at forward angles even
 larger value of $R$.
 We checked that the same picture of  approaching the screening limit
 is seen for all other elastic scattering
 spin observables (altogether 55 observables, including
 proton analyzing power, deuteron
 vector and tensor analyzing powers, spin correlation coefficients, as
 well as 
 spin transfer coefficients from the nucleon or deuteron to the nucleon
 or  deuteron). The achieved screening limit for a particular observable
 is equal to the prediction for that observable obtained with the limiting
 off-shell Coulomb amplitude of Eq.~(\ref{73a})
 (brown solid lines in Fig.~\ref{fig1}), what is a very strong test of
 reaching the screening limit.

\begin{figure}
\includegraphics[scale=0.6]{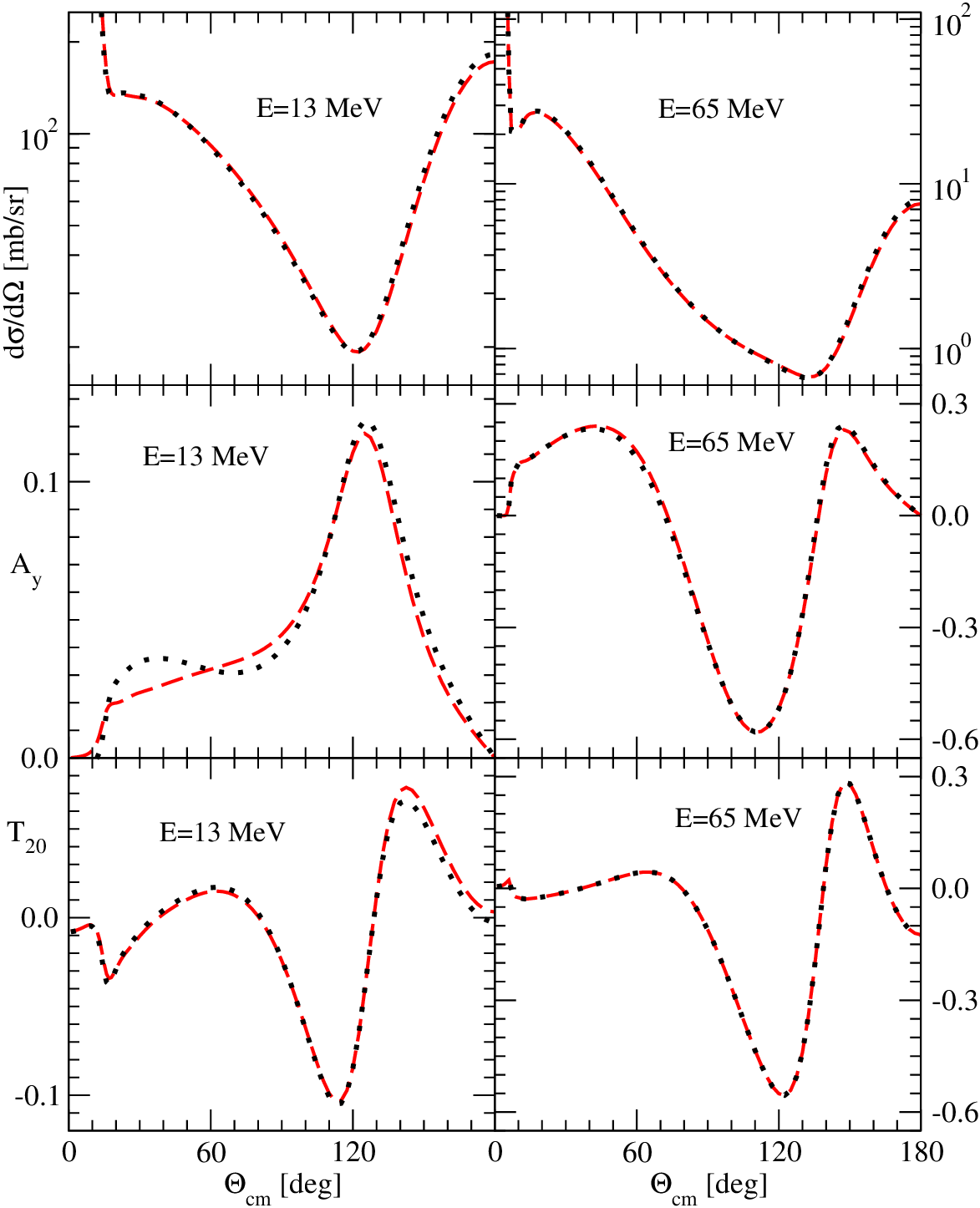}  
\caption{(color online)
 The predictions for the  pd elastic scattering cross section
 $\frac {d\sigma} {d\Omega}$, proton analyzing power $A_y$, and for
 the deuteron tensor analyzing power $T_{20}$, 
 shown as a function of the c.m. angle
$\Theta_{cm}$ at the incoming lab. proton energy
 $E=13$ and $65$~MeV.
 The black dotted lines show results of the approach based on 
 Eq.~(\ref{63}) 
 and the red dashed lines on Eq.~(\ref{65}).
 In both cases the elastic scattering
 transition amplitude (\ref{79}) was used.  
 These observables were
 calculated with the screened Coulomb force ($R=40$~fm, $n=4$)
 and the AV18 nucleon-nucleon potential~\cite{av18}
restricted to the $j \le 3$ partial waves.
}
 \label{fig2}
\end{figure}

 At $E=13$ and $65$~MeV we compared  results of the approaches based
 on Eqs.~(\ref{63})   
 and (\ref{65}), taking the limiting screening value of $R=40$~fm. 
 In Fig.~\ref{fig2} the  predictions
 for the cross section,  proton analyzing power $A_y$, and the deuteron
 tensor analyzing power $T_{20}$, of
 the first (Eq.~(\ref{63}) - the black dotted line)
 and second (Eq.~(\ref{65}) - the red dashed line) approach are
 shown.
 There is a nice agreement between predictions of both approaches
 at $E=65$~MeV, which
 extends also to other, not shown, observables. At $E=13$~MeV for
 some spin observables, like  $A_y$ shown in Fig.~\ref{fig2},
 differences appear, but generally also here the agreement
 is  good. These results demonstrate that even with such a rather small
 $j \le 3$ basis 
 cancellation effects cause that 
 our simplified approach works quite well. 

 The basic difference between these two approaches lies in the treatment of
 the first pair of contributing Coulomb terms. While the first approach  
 relies solely on  the cancellation of contributions of these terms, 
 in the second one they are calculated  explicitly.
 Since one expects that the magnitude of the Coulomb terms as well as of
 the Coulomb force effects  diminishes with increasing energy, the
 above mentioned differences at $13$~MeV could be
 interpreted as a warning that one should  avoid
 a direct calculation of complicated terms containing  3-dimensional
 Coulomb t-matrix and, instead,  
  rely on the cancellation between contributing Coulomb terms.
 Presently the terms with 3-dimensional Coulomb t-matrix can be calculated
 only for very restricted sets of partial wave states
 $\left | {\alpha } \right\rangle$ and total 3N angular momenta $J$,
  highly  insufficient in full-fledged calculations needed
 for analyses of data. 

 In the following we concentrate on the first approach showing how its
 precision can be improved and performance controlled.   
 It is evident that increasing the number of partial wave states
 $\left | {\alpha } \right\rangle$ would strengthen the cancellation
 effect between contributing Coulomb terms, 
 improving thus approximation (\ref{63}).
 Namely, for any particular value of the screening radius $R$ there is
 a finite number of partial wave states which reproduce exactly the
 3-dimensional Coulomb t-matrix. In the extreme case when
 $\left | {\alpha } \right\rangle$ is taken as such a set of states, 
 the Coulomb terms in the Faddeev equation
 (\ref{55}) cancel exactly and 
 the approach based on (\ref{63}) becomes an exact one.
 Otherwise it is only an approximation, quality of which depends
 on how large is the cancellation effect between Coulomb terms.
 Since that cancellation concerns not only Coulomb terms in Eq.~(\ref{63})
 but also to some extent in elastic scattering and breakup
 transition amplitudes of Eq.~(\ref{78}) ( (\ref{79}) ),
 the condition reflecting degree
 of cancellation can be made
 quantitative by comparing cross sections (observables)  obtained
 with only the first term and with all terms in (\ref{78}) ( (\ref{79}) ).
 In  the case of a complete cancellation they should be
 equal. However, in many cases it will be sufficient when these two results
  converge with an increasing basis $\left | {\alpha } \right\rangle$ 
 even to  different values.
 Starting from an initial set of $\left | {\alpha } \right\rangle$-states
 one needs to extend it by incorporating   consecutive
 states from $\left | {\beta } \right\rangle$.

 To study this in more detail we extended the 
 set  of $\left | {\alpha } \right\rangle$-states by adding to the 
 initially chosen states  with $j \le j_s=3$, in which
both strong interactions and pp Coulomb force  act, some 
partial wave states from set 
   $\left | {\beta } \right\rangle$ 
with  $j_s \le  j \le j_{max}$, in which only pp Coulomb
force operates. In the following such an extended  set of
$\left | {\alpha } \right\rangle$-states
 will be denoted by ``$jsj_sjj_{max}$'',  so that the initially  
 used set is denoted by $js3j3$. We solved  Eq.~(\ref{63})
 at $E=10$~MeV for
a number of extended   $\left | {\alpha } \right\rangle$ sets:
$js3j5, js3j7, js3j8, js3j9$, and $js3j10$, and
looked for a pattern of convergence for different elastic scattering
observables with a growing number
of $j_{max}$. The increase in number of treated  partial waves for
given total angular momentum  $J$ and parity $\pi$ of the $ppn$ system is
 large and amounts to $89$ for $js3j3$, 
 $165$ for $js3j5$, and  $539$ for $js3j10$. 
In spite of that increase the time required to solve numerically Faddeev
equations remains restricted due to the fact that partial waves of pure
Coulomb nature $\left | {\beta } \right\rangle$  
(with $j_s \le j \le j_{max}$)  do not couple between themselves
(see a remark about the third term after (\ref{8a})
and Eq.~(2) in Ref.~\cite{delt2006br}). 

We found for  practically all elastic scattering observables that
the convergence
in $j_{max}$ is rapid. The most influenced by changes of $j_{max}$
 are low energy proton and deuteron vector analyzing powers, $A_y$
 and $iT_{11}$,  which require for converged result at $10$~MeV
 the basis $js3j7$.
 At $65$~MeV and higher energies it is sufficient to use the basis $js3j3$,
 what reflects the diminishing  pp Coulomb force effects
 for higher energies.  
 
\begin{figure}
 \includegraphics[scale=0.6]{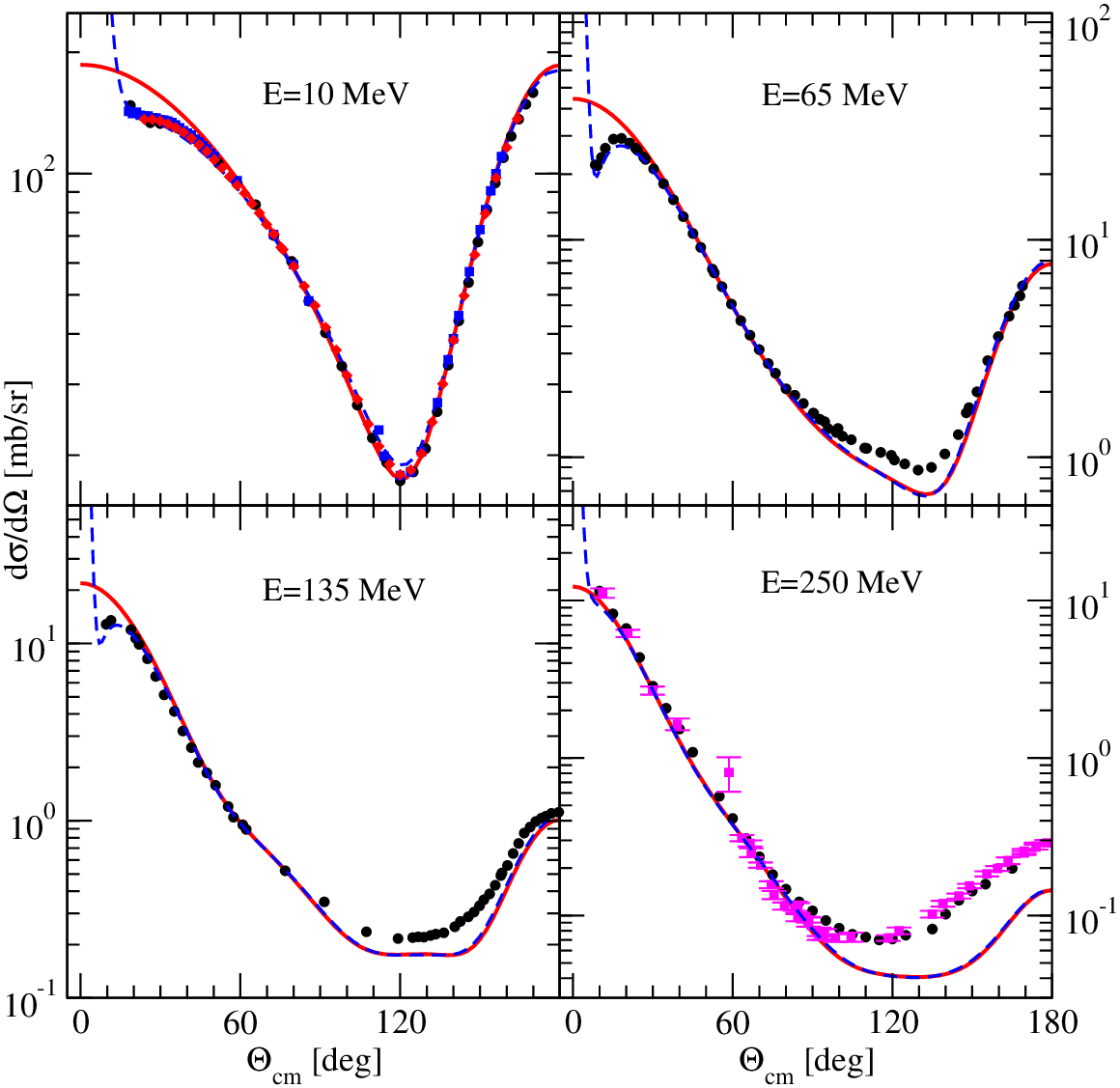} 
\caption{
  Comparison of  data and predictions for the
  pd elastic scattering cross section
$\frac {d\sigma} {d\Omega}$ obtained in the approach based on 
 Eq.~(\ref{63}) and using elastic scattering
transition amplitude (\ref{79}),  at the incoming proton lab. energy
$E=10, 65, 135,$ and $250$~MeV. The  cross sections were
calculated with the screened Coulomb
force ($R=40$~fm, $n=4$) and the AV18 nucleon-nucleon potential~\cite{av18}
restricted to the $j \le 3$ partial waves. 
 The pure Coulomb term
 $\left\langle {\Phi' } | Pt_cP| {\Phi } \right\rangle$
 was determined according to Eq.~(\ref{73a})  
 (blue dashed line).
 At $E=10$~MeV the set $js3j7$ of
 $ \left| {\alpha } \right\rangle$-states was used while for other
 energies the set  $js3j3$. 
 The red solid line is the corresponding
 nd elastic scattering cross section.
 The black circles, blue squares and red diamonds at  $E=10$~MeV
 are  pd elastic scattering cross section 
 data of Ref.~\cite{grub12}, ~\cite{schieck}, and ~\cite{sagara},
 respectively.
 The black circles at $E=65$~MeV are pd data from ~\cite{shimizu}, at
 $E=135$~MeV  from  ~\cite{sak99},  and at $E=250$~MeV from ~\cite{hatanaka}.
  The magenta squares at $E=250$~MeV are nd data from ~\cite{maeda250}. 
 }
 \label{fig3}
\end{figure}

\begin{figure}
\includegraphics[scale=0.6]{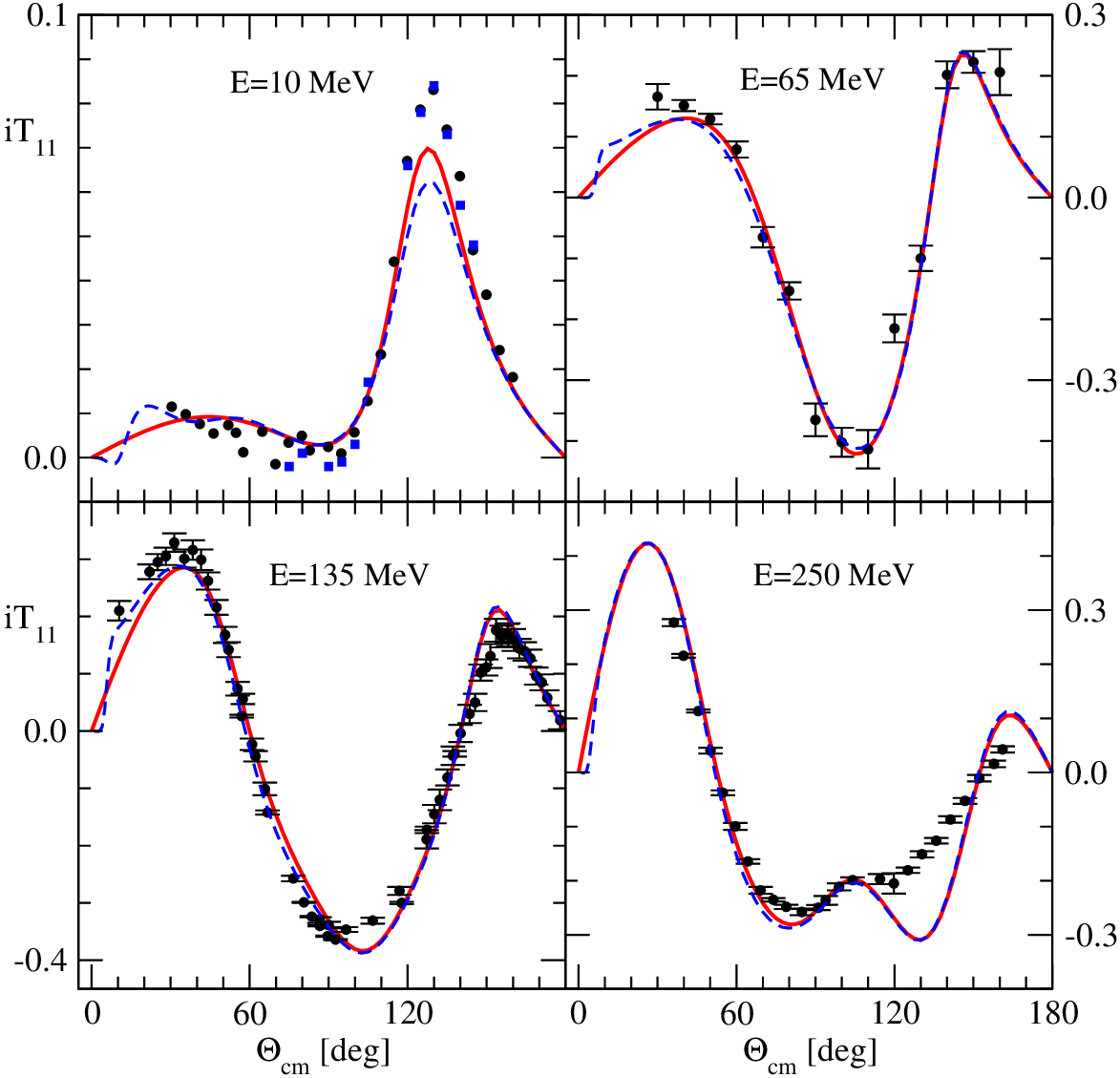}  
\caption{(color online) 
  The same as in Fig.~\ref{fig3} but for the deuteron vector analyzing power
  $iT_{11}$. For the description of lines see Fig.~\ref{fig3}.
  The black circles are pd data from ~\cite{sperisen84} at  $E=10$~MeV,
  ~\cite{cameron} at  $E=65$~MeV, ~\cite{seki140} at  $E=135$~MeV, and
  ~\cite{seki250} at  $E=250$~MeV. 
  The blue squares at  $E=10$~MeV are pd data from ~\cite{sawada83}.
}
 \label{fig4}
\end{figure}

\begin{figure}
\includegraphics[scale=0.6]{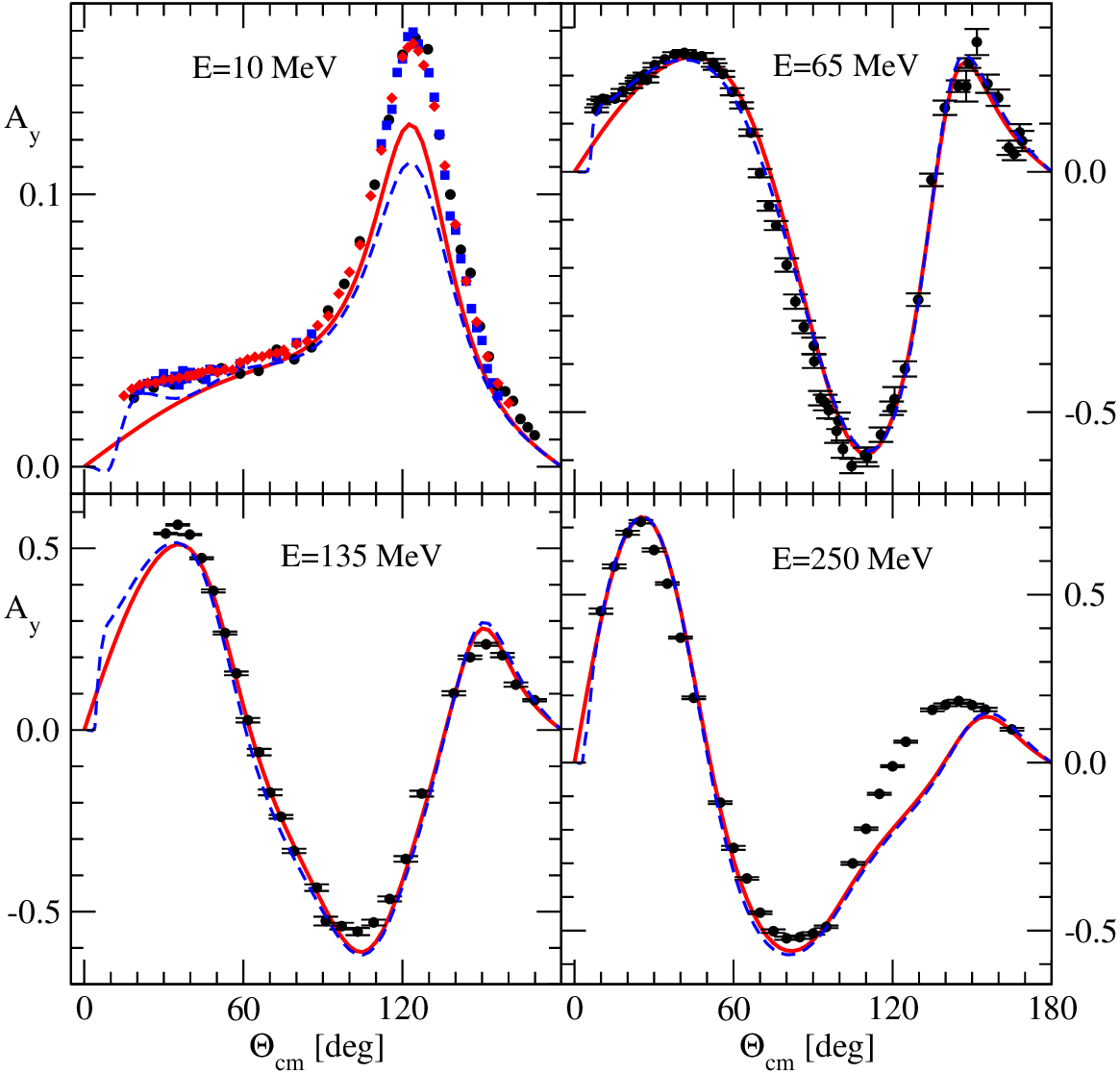}  
\caption{(color online)
  The same as in Fig.~\ref{fig3} but for the proton analyzing power
  $A_{y}$. For the description of lines see Fig.~\ref{fig3}.
  The black circles are pd data from ~\cite{sperisen84} at  $E=10$~MeV,
  ~\cite{shimizu} at  $E=65$~MeV, ~\cite{ermisch} at  $E=135$~MeV, and
  ~\cite{hatanaka} at  $E=250$~MeV. 
  The blue squares at  $E=10$~MeV are pd data from ~\cite{schieck} and
  red diamonds from ~\cite{sagara}.  
}
\label{fig5}
\end{figure}

\begin{figure}
\includegraphics[scale=0.6]{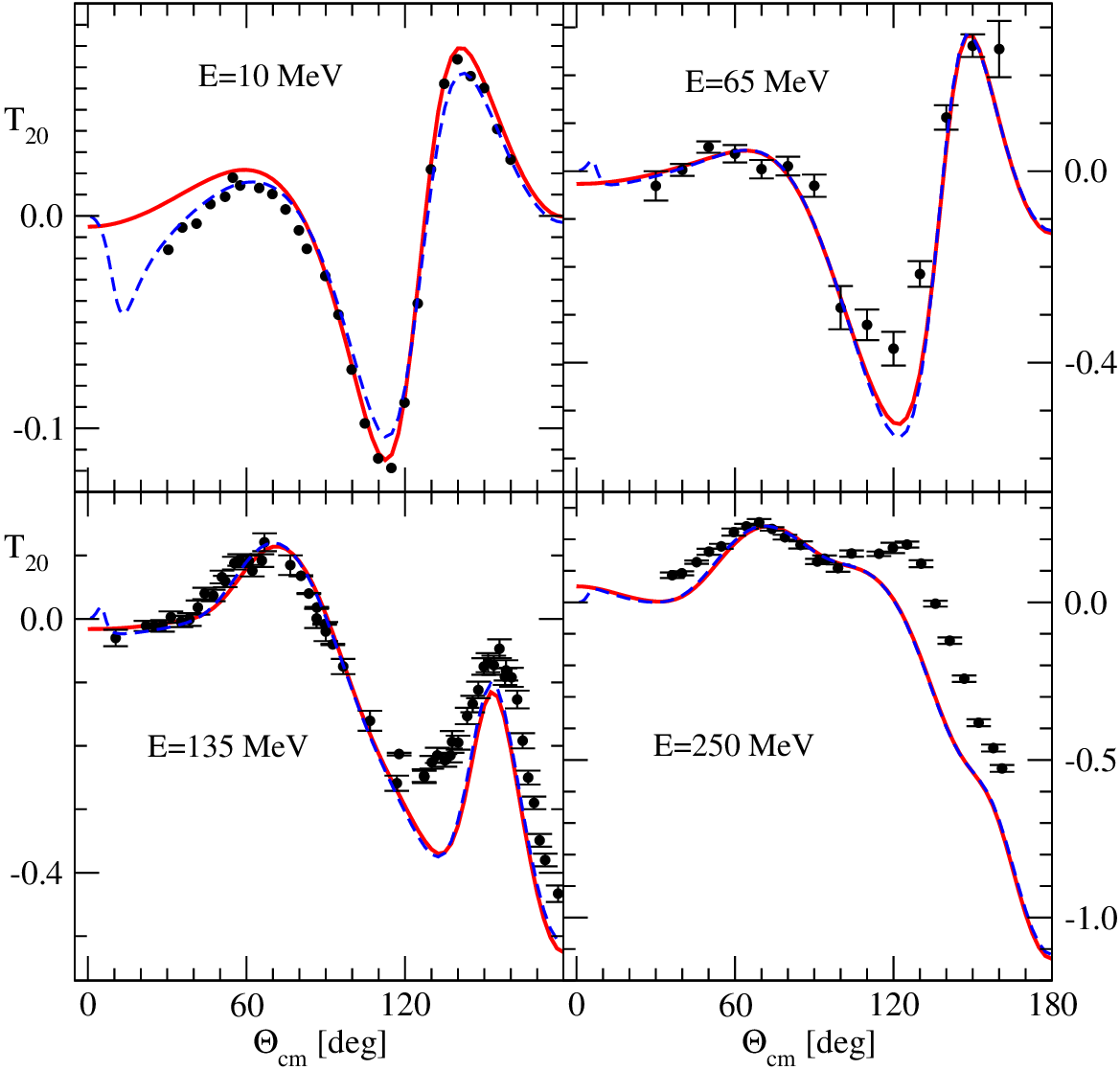}  
\caption{(color online)
  The same as in Fig.~\ref{fig3} but for the deuteron tensor analyzing power
  $T_{20}$. For the description of lines see Fig.~\ref{fig3}.
  The black circles are pd data from ~\cite{sawada83} at  $E=10$~MeV,
  ~\cite{cameron} at  $E=65$~MeV, ~\cite{seki140} at  $E=135$~MeV, and
  ~\cite{seki250} at  $E=250$~MeV. 
}
\label{fig6}
\end{figure}

 In Figs.~\ref{fig3}-\ref{fig6} we compare at different energies
 pd predictions of the
 first approach and using elastic scattering amplitude (\ref{78})
 containing only the first pair of Coulomb terms   
 (taking the limiting screening value of $R=40$~fm and
 calculating the Coulomb term
$\left\langle {\Phi}' \right| Pt_c^{R}P\left| {\Phi} \right\rangle$ 
 with the limiting
 off-shell Coulomb t-matrix of Eq.~(\ref{73a})), 
  to available pd data for the elastic scattering cross
 section (Fig.~\ref{fig3}), the deuteron vector analyzing power $iT_{11}$
 (Fig.~\ref{fig4}), the proton analyzing power $A_y$ (Fig.~\ref{fig5}), and
 the deuteron tensor analyzing power $T_{20}$
 (Fig.~\ref{fig6}). To emphasise the  importance and magnitude of Coulomb
 effects   we provide also nd predictions for these observables.
 Large pp Coulomb force effects for elastic scattering observables
 are concentrated mainly at forward angles
 below $\theta_{c.m.} \approx 30^o$ and they
 diminish with an increasing energy. For the cross section the characteristic
 pattern caused by the pp Coulomb force is properly reproduced
 by the calculations.
 Also the forward angle data for $A_y$, $iT_{11}$, and $T_{20}$
  are nicely reproduced. It follows that large discrepancies
 between pd cross section data  and predictions at middle and backward
 c.m. angles, which grow with the increasing
 energy, are not caused by the pp Coulomb force and must be explained either
 by the action of three-nucleon forces (3NF's) ($E=65$ and
 $135$~MeV) ~\cite{epel2020}
 or through activation of mesonic degrees of freedom
 ($E=250$~MeV) ~\cite{wit2022}.

 The above results were obtained omitting completely the second pair of
 Coulomb terms, namely the third and fifth term, 
 in the elastic scattering transition amplitude (\ref{78}).
 To answer the question, how are the elastic scattering
 observables affected by approximation (\ref{79}) for the scattering
 amplitude, numerical calculation of both neglected terms are required. 
 The computation of the first term in the second pair 
 $$-\left\langle {\vec p\vec q~} \right|\sum\limits_{\alpha '} {\int
{p'^2 dp'q'^2 dq'\left| {p'q'\alpha '} \right\rangle \left\langle
{p'q'\alpha '} \right|} } t_c^R PG_0
 \sum\limits_{\alpha
''} {\int {p''^2 dp''q''^2 dq''\left| {p''q''\alpha ''}
\right\rangle \left\langle {p''q''\alpha ''} \right|} } T\left|
            {\Phi } \right\rangle$$ 
is straightforward but determination of the second one 
 $$ \left\langle
{\vec p\vec q~} \right|t_c^R PG_0 \sum\limits_{\alpha '} {\int
{p'^2 dp'q'^2 dq'\left| {p'q'\alpha '} \right\rangle \left\langle
  {p'q'\alpha '} \right|} } T\left| {\Phi } \right\rangle , $$ 
which contains the 3-dimensional Coulomb t-matrix $t_c^R$, 
presents quite a formidable numerical task according to
expression (D.9), (D.6), and (D.8) of Ref.~\cite{elascoul}.
We postpone direct numerical calculation of this term till a future
study and here we would like to present some plausible
arguments which justify omission of both terms in  (\ref{78}).
To that end we investigated changes of elastic scattering observables
induced by inclusion of the first term in the calculation of observables
at our four energies with $js3j3$ set
$\left| {\alpha }  \right\rangle$. It turned out that at
    $E=65, 135$, and $250$~MeV, the modifications are practically
    negligible for all $55$ elastic scattering observables. At the lowest
    investigated energy $E=10$~MeV, some spin observables were modified by
    $\approx 5 - 10~ \%$. Changing the set $js3j3$ to $js3j7$ led to the
    similar picture at $10$~MeV. It permits us to conclude that the
    magnitude of the first term diminishes with the increasing energy and at
    the higher energies this term can be safely dropped.
    At the low energies its contribution is not overwhelmingly large.
    That together
    with the fact that also in elastic scattering amplitude (\ref{78}) one
     expects a cancellation between contributions from the second pair
    of the Coulomb terms similar to that for the first pair, 
    seems to justify the omission of the  two terms discussed.

\begin{figure}
\includegraphics[scale=0.5]{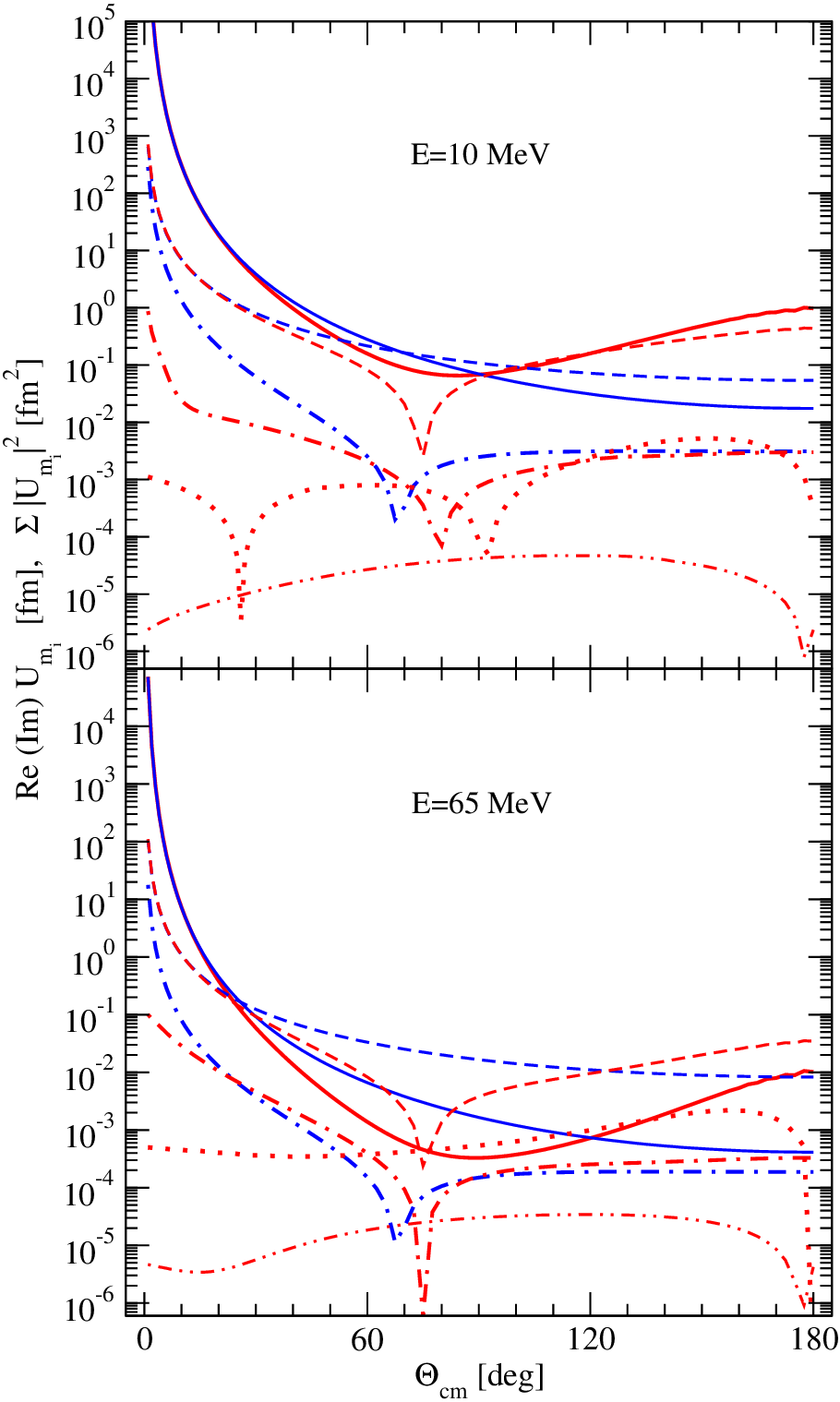}  
\caption{(color online) 
  Comparison of  the extended deuteron
   $\left\langle {\Phi' } | Pt_cP| {\Phi } \right\rangle$
  (Eq.~(\ref{73a}))
  and the  point deuteron $A_c$~\cite{haer1985} pd scattering
  Coulomb amplitudes.
  The red and blue solid lines are sums, over all incoming
  and outgoing proton and deuteron spin projections, of the squares of
  the extended and point deuteron amplitudes, respectively. The blue 
  short dashed and dashed-dotted lines
  (the red short-dashed and dashed-dotted)
  are absolute values of real (imaginary) parts of  
  the point  and extended deuteron amplitudes, respectively,
  for transition from
  the incoming proton
  ($-\frac {1} {2}$) and deuteron ($-1$) spin projections to the outgoing
  ($-\frac {1} {2}$) and  ($-1$) ones.
  The red dotted  and red  dashed-double-dotted lines 
  are absolute values of real and imaginary parts of  
  the extended deuteron  amplitude, respectively,
  for nondiagonal transition from
  the incoming proton
  ($-\frac {1} {2}$) and deuteron ($0$) spin projections to the outgoing
  ($-\frac {1} {2}$) and  ($-1$) ones. 
}
\label{fig7}
\end{figure}

 Last but not least,  
 in Fig.~\ref{fig7} we compare at two lab. energies of the incoming proton
 $E=10$ and $65$~MeV the point deuteron 
 pd Coulomb amplitude $A_c$ ~\cite{haer1985} with the extended deuteron one, 
 $\left\langle {\Phi } | Pt_cR| {\Phi } \right\rangle$, 
 calculated with the limiting Coulomb t-matrix of Eq.~(\ref{73a}). 
 While the $A_c$ is diagonal in incoming
 and outgoing proton and deuteron spin projections, the extended deuteron
 amplitude permits  also nondiagonal terms, which however are few orders
 of magnitude smaller than
 the diagonal ones. The red and blue solid  lines
 are sums, over all incoming
  and outgoing proton and deuteron spin projections, of the squares of
  the extended and point deuteron amplitudes, respectively. 
  There is a nice agreement between both amplitudes in the most
  important region of angles up
  to $\approx 30^o$, not only for that sum but also for the magnitudes
  of their real parts.
The imaginary parts are much more smaller.  It is illustrated in Fig~\ref{fig7}
for the case of a transition from $m_d^{in}=-1,m_p^{in}=-\frac {1} {2}$ to
  $m_d^{out}=-1,m_p^{out}=-\frac {1} {2}$. Also the  non diagonal term
    for the extended deuteron,  for transition from    
 $m_d^{in}=0,m_p^{in}=-\frac {1} {2}$ to
    $m_d^{out}=-1,m_p^{out}=-\frac {1} {2}$, is shown.

\subsection{Breakup reaction}
\label{breakup}

The exclusive breakup reaction offers  a rich
spectrum of kinematically complete geometries with  observables 
 sensitive to underlying dynamics. 
We decided to focus on three  geometries specified by a kinematical
condition for momenta of the three outgoing nucleons.
 In the so called final-state-interaction (FSI)  geometry the two outgoing 
 nucleons have equal momenta. In neutron-deuteron (nd) breakup
 their strong interaction in the $^1S_0$ state leads to a 
characteristic cross section maximum occurring at the exact FSI condition,  
the magnitude of which is sensitive to the $^1S_0$ scattering length.
In the symmetrical-space-star (SST) configuration
the momenta of the three outgoing nucleons in the 3N center-of-mass (c.m.)
have the same 
magnitudes and  form a three-pointed "Mercedes-Benz" 
star. That star  lies in a plane inclined under  an angle $\alpha$ 
with respect to the beam direction with momenta of the two outgoing
and detected nucleons (in our case protons) lying symmetrically to the beam. 
 The quasi-free-scattering (QFS) geometry  refers to a situation
 where one of the nucleons is at rest in the
laboratory system. In pd breakup the np or pp QFS configurations are
possible, while for nd breakup np or neutron-neutron (nn) quasi-freely
scattered pairs can emerge.

The characteristic feature of the exact approach  of Refs. ~\cite{elascoul}
and ~\cite{brcoul} as well as of our simplified one 
is the appearance in the breakup transition amplitude 
 of a new term, $\left\langle \vec p_0 \vec q \right| (1+P)t_c^RP 
 \left| {\Phi} \right\rangle$,
 based on a 3-dimensional Coulomb t-matrix, $t_c^R$, 
 analogous to the
 Rurherford term in the elastic pd scattering. The expression for
 this term given in Appendix C of Ref.~\cite{elascoul} (Eq.~(C.3)) shows that
 largest contributions from this term 
 are expected in the region of the breakup phase space where the argument of
 the deuteron wave function,  
 ${\varphi_L(\vert \vec q + \frac {1} {2} \vec q_0 \vert )}$, vanishes. That  
 condition  $\vec q = - \frac {1} {2} \vec q_0$  occurs in the pp
 QFS, where the spectator neutron rests in the  laboratory system 
 (see also the discussion on p.185 of Ref.~\cite{physrep96}). Therefore one 
 expects large pp Coulomb force effects for that geometry. 
 Also  large Coulomb effects are
 expected in the pp FSI region,
 where two outgoing protons have equal momenta and interact strongly.
 For nn FSI this leads to a pronounced  cross section maximum just at
 the nn FSI condition and in the case of pp FSI the Coulomb barrier should 
 prevent such a maximum from forming.  

 In the following we start to investigate the breakup reaction using approach
 (\ref{63}) with the set $js3j3$ and the breakup transition amplitude
 of Eq.~(\ref{79}). 
 First we demonstrate the pattern of convergence to the screening
 limit  in  two  exclusive geometries, QFS as well as FSI, and  show that
 the final results do not depend  on how specifically the on-shell
 breakup amplitudes, which  undergo  renormalization, are derived. 
 We will apply renormalization to the on-shell breakup
 amplitudes obtained in two different ways. 
 In the first approach the on-shell breakup amplitudes
 $\left\langle p_0  q \alpha \right| T^R \left| {\Phi} \right\rangle$ 
 ($\left\langle p_0  q \alpha \right| t_c^RP \left| {\Phi} \right\rangle$)
 are obtained by interpolation from   the off-shell ones
 $\left\langle p  q \alpha \right| T^R \left| {\Phi} \right\rangle$  
  ($\left\langle  p  q \alpha \right| t_c^RP \left| {\Phi} \right\rangle$), with
 subsequent removal of the oscillating phase factor $e^{i\Phi^R_n}$ when
  calculating the breakup transition amplitude.
  In the second method we generate  the half-shell pp t-matrix
  $t_{N+c}^R(p_0,p';E-\frac{3} {4m} q^2)$
  ($t_c^R(p_0,p';E-\frac{3} {4m} q^2)$)  
  and calculate the 
  on-shell transition matrix elements 
  $\left\langle p_0  q \alpha \right| T^R \left| {\Phi} \right\rangle$
  ($\left\langle p_0  q \alpha \right| t_c^RP \left| {\Phi} \right\rangle$) 
  according to Eq.~(\ref{eq80}). 
  Here  one has to use unrenormalized $t_{N+c}^R(p_0,p';E-\frac{3} {4m} q^2)$
 ($t_c^R(p_0,p';E-\frac{3} {4m} q^2)$)  and postpone again the renormalization
  to calculation of the breakup  transition  amplitude. 
 
\begin{figure}
\includegraphics[scale=0.6]{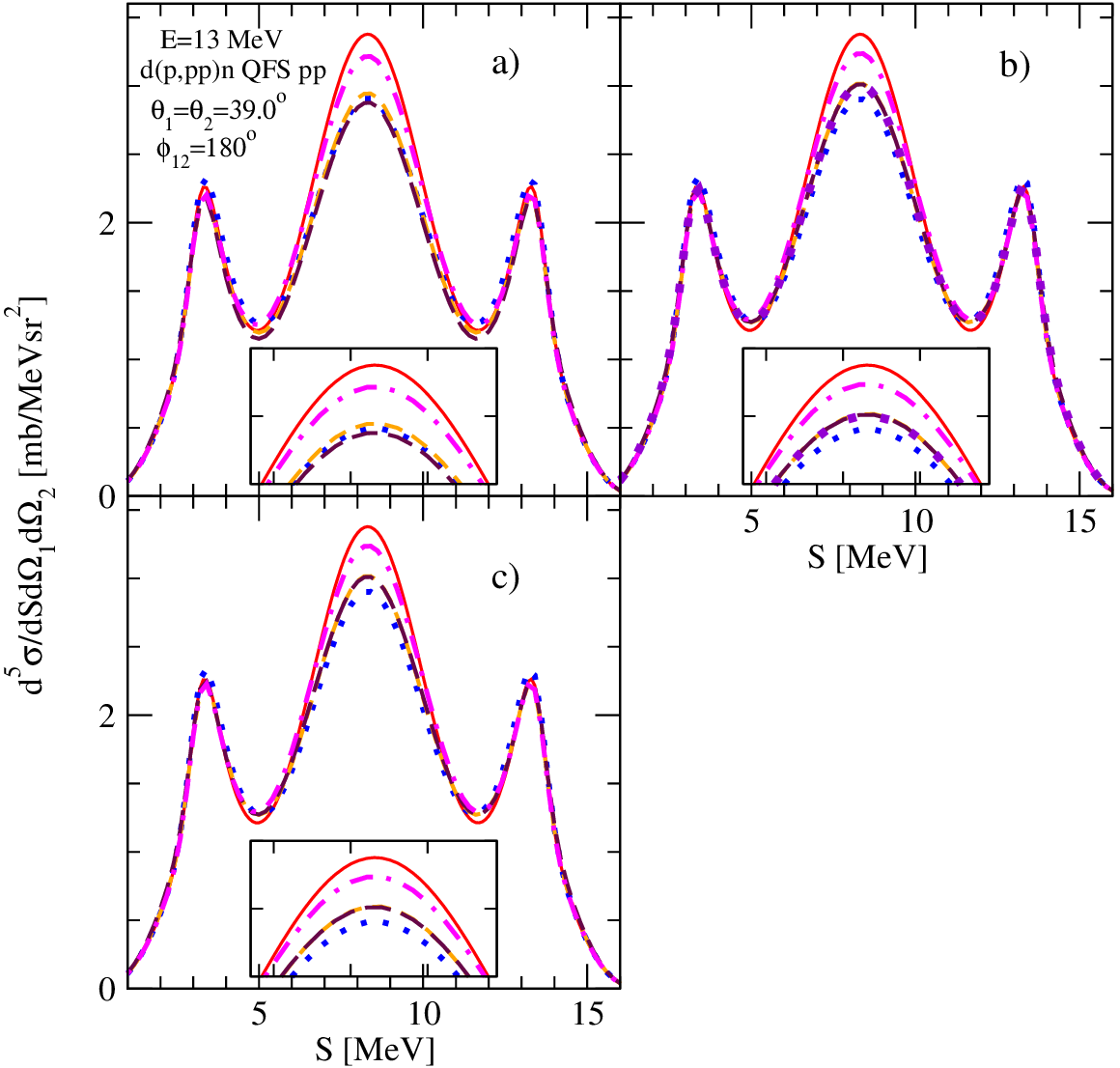}  
\caption{(color online) The convergence with respect to the cut-off radius R
of the  pd breakup cross section
$\frac {d^5\sigma} {d\Omega_1d\Omega_2dS}$ shown as a function of
the S-curve length for the pp QFS kinematically complete configuration
at the incoming proton lab. energy
$E=13$~MeV.  
 The screening
 radii are  ($n=4$):  $R=5$~fm (blue dotted line),
 $R=10$~fm (magenta dashed-dotted line), 
 $R=20$~fm (orange short-dashed line),
 $R=40$~fm (maroon long-dashed line). 
 These cross sections were
calculated   using the approach based on Eq.~(\ref{63}) 
   with the screened Coulomb
force and the AV18 nucleon-nucleon potential~\cite{av18}
restricted to the $j \le 3$ partial waves (set $js3j3$), taking the on-shell
Faddeev amplitudes obtained in two  different ways
and applying renormalization when calculating the breakup transition amplitude.
  In a) unrenormalized
on-shell amplitudes gained 
by interpolation from the off-shell ones were used. In b) the on-shell
amplitudes of a) have been renormalized before calculating observables.
In c) the on-shell amplitudes were calculated according to (\ref{eq80}) with
 the unrenormalized pp part of $t_{N + c}^R$ ($t_c^R$) 
 and renormalization was performed
 before calculating observables.
 The red solid line is the corresponding
nd elastic scattering cross section. The violet dotted line in b)
 is the result with $R=40$~fm but the pure Coulomb term
 $\left\langle {\vec p_0 \vec q} | (1+P)t_cP| {\Phi } \right\rangle$
  determined with the $t_c$ of Eq.~(\ref{73b}).   
 }
 \label{fig8}
\end{figure}

  In Fig.~\ref{fig8} we present the pattern of convergence in the
  screening radius $R$ for
  the pp QFS. Taking unrenormalized on-shell breakup amplitudes obtained by an 
  interpolation from the off-shell solutions of the Faddeev equations to
  the on-shell  values $(p_0,q)$ provides unrenormalized pp QFS
  cross sections which
  change with varying $R$ (Fig.~\ref{fig8}a). Renormalizing these amplitudes
  stabilizes the cross sections for the screening radii $R \ge 20$~fm
  (see Fig.~\ref{fig8}b). The limiting values of the pp QFS cross sections
  do not depend on the way the on-shell breakup amplitudes
  are determined (see Figs.~\ref{fig8}b and
  \ref{fig8}c).  These two 
  methods lead to the same final pp QFS
  cross sections. That the screening limit has been achieved is confirmed 
  in Fig.~\ref{fig8}b, where the violet dotted line 
 shows the result for $R=40$~fm with the pure Coulomb term
 $\left\langle {\vec p_0 \vec q} | (1+P)t_cP| {\Phi } \right\rangle$
 determined using the final  3-dimensional Coulomb t-matrix 
 of Eq.~(\ref{73b}). 

\begin{figure}
\includegraphics[scale=0.6]{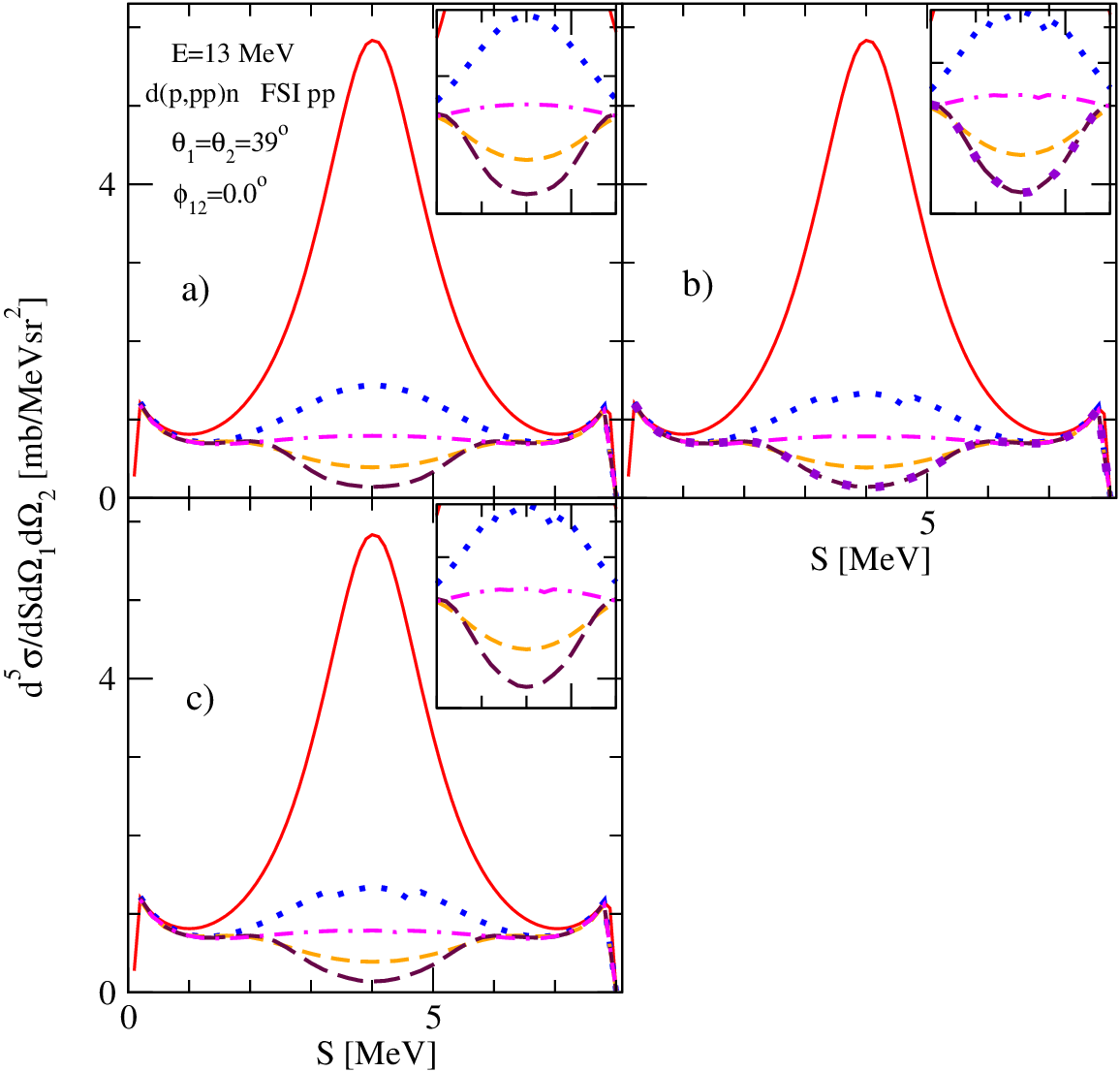}  
\caption{(color online)
  The same as in Fig.~\ref{fig8} but for the pp FSI kinematically complete
  configuration.
}
 \label{fig9}
\end{figure}

  In Fig.~\ref{fig9} we present analogous investigation for the pp FSI
  configuration. The large effect of the pp Coulomb force is seen
  in the region of FSI, where,  
  instead of a clear maximum present for nn FSI, the cross section
  is reduced practically to zero by the pp Coulomb barrier. The pattern of
  approaching  the limiting value is similar as in the
  case of the pp QFS and final result also here 
  does not depend on how the on-shell breakup amplitudes were derived. 
  To get the
  final values of the cross section one needs to go to a larger 
  screening radius than in the case of the pp QFS (see Fig.\ref{fig9}b).

\begin{figure}
\includegraphics[scale=0.49]{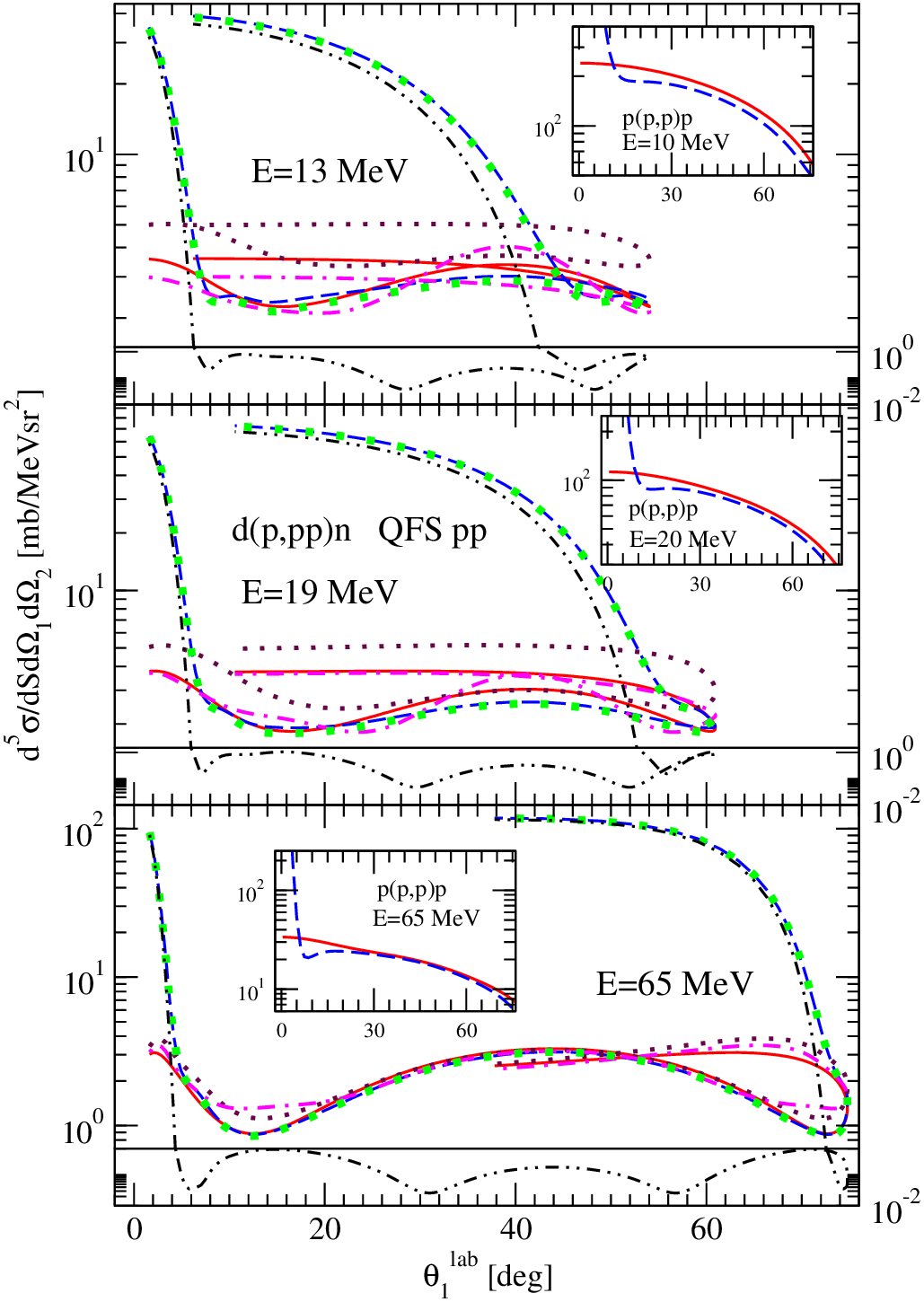}  
\caption{(color online)
The  pd  breakup d(p,pp)n 
 cross section calculated  at 
 the pp QFS condition   
for the incoming proton lab. energy $E=13, 19,$ and $65$~MeV,  
  as a function of the  
 angle  of the outgoing proton 1. 
 The approach based on Eq.~(\ref{63})
 with set of $js3j3$ partial waves and the screening radius $R=40$~fm ($n=4$)
 was used, 
 with on-shell
 Faddeev amplitudes obtained by an interpolation from the off-shell ones,
 renormalized before calculating observables.
 The  blue long-dashed line
 is the result with all three terms in Eq.~(\ref{79}).
 Also the result without 
 renormalization is shown by the green dotted line.  
 The maroon dotted  and magenta dashed-dotted lines follow when
 the  term with 3-dimensional $t_c^R$ Coulomb t-matrix and both terms with
 $t_c^R$ are omitted, respectively, and the black double-dotted-dashed
 line when only these two terms are kept.
 At bottom of each figure continuation of the black double-dotted-dashed line
 is shown in a compressed y-axis scale  shown on the right side. 
 The red solid line is the nd breakup cross section.
  In insets the lab. cross sections
$d\sigma/d\Omega$ ($\frac {mb} {sr}$) for pp scattering at lab. 
energies $E=10, 20,$ and $65$~MeV are shown as a function of the
 proton lab. angle. Here the red solid and blue dashed lines
are the AV18 cross sections without and with pp Coulomb
interaction, respectively. 
 }
 \label{fig10}
\end{figure}

  In Fig.~\ref{fig8} non-negligible  effects of renormalization
  (Fig.~\ref{fig8}a and b),  
  which raise the cross section by $\approx 5 \%$ in the maximum,
  are seen for that
  particular pp QFS configuration. In order to investigate the
  renormalization 
  as well as  the pp Coulomb force effects  
  for all pp QFS configurations, we looked at the pp
   quasi-free-scattering cross section exactly
  at the QFS condition (maximum of the cross section) as a function of the lab.
  angle of the first outgoing proton  $\theta_1^{lab}$ ($d(p,p_1p_2)n$).
  Since later we will compare theoretical predictions to available
  pp QFS cross section  data at $E=9.5, 13, 19, 22.7,$ and $65$~MeV, we show 
  in Fig.~\ref{fig10} results of this investigation for
  three energies: $E=13, 19,$ and $65$~MeV. 
  At each energy at given $\theta_1^{lab}$ there are 2 solutions for
  the QFS condition,
  the second one (upper branch in Fig.~\ref{fig10}) being not (or difficult) 
  accessible to 
  measurement due to too small energy of one of the outgoing protons.
  The full result with renormalization is given by the
  blue long-dashed line which compared with the green dotted 
  line (without renormalization) reveals importance and the magnitude of
  renormalization. The non-negligible  renormalization effects
  at $E=13$~MeV of the order of $\approx 5-8 \%$ are seen
  only for $\theta_1^{lab} \in (10^o,50^o)$.
  At $19$ and $65$~MeV the renormalization
  causes in pp QFS only insignificant effects.

  In order to
  get the information of the Coulomb force effects in Fig.~\ref{fig10}
  also the nd prediction is shown by
  the solid red line. 
  The Coulomb effects are largest at $E=13$~MeV
  as evidenced by the nd results as well as by the maroon dotted line, 
  resulting when the term with the 3-dimensional Coulomb t-matrix
  is omitted, or
  by magenta dashed-dotted line, when both Coulomb terms are absent.
  They diminish rapidly with growing energy,
  becoming practically  negligible at $65$~MeV. 
   It is
  interesting to note that the angular dependence of the pp QFS cross section
  resembles that of pd elastic scattering, with the
   characteristic increase at
  forward angles caused by the Coulomb term
  $\left\langle \vec p_0 \vec q \right| (1+P)t_c^RP 
 \left| {\Phi} \right\rangle$. Also the importance and magnitude of
  the Coulomb force effects in pp QFS resembles those
  in a free pp scattering (see insets in Fig.~\ref{fig10}).
  To exemplify the cancellation effect between the first pair of
  Coulomb terms in the pp QFS
  breakup transition amplitude we show in Fig.~\ref{fig10} also
   cross sections (black double-dotted-dashed line) obtained with
   these two terms only. In the region of angles
   $\theta_1^{lab} \in (10^o,40^o)$ 
  the resulting values of that cross section are
  about $\approx 10^{-1} \frac {mb} {MeV*sr^2}$
  for that set of $| \alpha >$-states, what illustrates quite significant  
  cancellation, 
   however not so large  as for the case of the pp FSI
    or pd SST (see below).

    In Figs.~\ref{fig11} and \ref{fig12}  results of a similar investigation
    are shown for pp FSI and pd SST, respectively.
    The action of the Coulomb barrier brings the pp FSI
  cross section close to zero for all pp FSI configurations.
  Regardless from their production angle
  $\theta_1^{lab}$, the renormalization effect is insignificant and
  the cancellation between two contributing Coulomb terms appears drastic
  (black double-dotted-dashed line). The pp FSI
  cross section is determined practically by only the first term
  in the breakup
  transition amplitude (\ref{79}). 
  For the SST configurations the largest effects of renormalization 
   as well as of the pp Coulomb force are present at $E =13$ and $19$~MeV
  around $\alpha_{cm} \approx 40^o$. They become again negligible at
  $E = 65$~MeV.  When the star plane is perpendicular to
  the beam direction the renormalization effects are small.
  As for the pp FSI, the cancellation effects in
  the breakup transition amplitude
  for that configuration are very large.

  In Figs.~\ref{fig13} and \ref{fig14} we show examples of comparison of our 
   theoretical predictions to pd
   breakup cross section data at $E=13$ and $65$~MeV for
   SST ($\alpha_{cm}=90^o$)
   and pp QFS  configurations. We show predictions of our two approaches, 
   one based on Eq.~(\ref{63}) 
   (blue short-dashed line) and second on  Eq.~(\ref{65}) 
   (maroon dotted line).  At $13$~MeV they agree very well with each other
   for pp QFS, 
   differing by $\approx 10 \%$  for SST.
   Both disagree significantly with the pd SST cross section data,
   underestimating  slightly  pp QFS data. At $65$~MeV they are close to each 
   other and  agree quite well  with the SST data,
   differing again only slightly from the pp QFS data.
   This is similar to what we have found for elastic scattering when comparing
   these two approaches and seems to support the validity of the assumption
   about the cancellation of the Coulomb terms  in the first approach, 
   at the same time indicating that  using directly computed 
   complex terms with 3-dimensional Coulomb t-matrix is hazardous.

   In  Ref.~\cite{sagara_qfs} a correct description
   of the low energy
   pp QFS cross section data was reported. 
   That prompted us to reanalyze  available low energy pp QFS
   cross section data and look for a possible reason of the slight discrepancy
   found at $13$~MeV in Fig.~\ref{fig13}.  
   In Fig.~\ref{fig15} we compare $E=19$~MeV data of Ref.~\cite{exp3} and
   $E=22.7$~MeV data of Ref.~\cite{miljanic} to our theoretical predictions
   obtained with the renormalized and unrenormalized $js3j3$ 
   on-shell breakup amplitudes of Eq.(\ref{79}),  
   received by interpolation from the off-shell ones (blue short-dashed and
   green dotted lines, respectively). The same comparison
   at $E=9.5$ and $13$~MeV for the data from Ref.~\cite{sagara_qfs} and
   at $13$~MeV for the data from Ref.~\cite{exp2}, 
   is presented  in Fig.~\ref{fig16}. 
   To make sure that the screening limit have been achieved we show in both
   figures by  the red dotted lines 
   also the renormalized $R = 40$~fm results obtained  with
   the pure Coulomb term calculated according to (\ref{73b}).

   A glance at both figures reveals again that the renormalization  effects
   shrink with the growing energy. 
   While at $9.5$ and $13$~MeV they are non-negligible 
   and enhance the cross section  bringing the theory closer to the data,
   at $19$ and $22.7$~MeV they are practically insignificant.
   It is interesting to notice that large effects of the pp
   Coulomb force at smaller energies have a pattern of contributions
   changing with pp QFS configuration (see the orange dashed-dotted and
   magenta double-dashed-dotted lines in Fig.~\ref{fig16}, which show the 
   results when both Coulomb terms are omitted). 
   At $9.5$ and $13$~MeV theory slightly underestimates pp QFS cross
   sections in practically all configurations.
   At $19$~MeV theory lies between two
   available data sets and at $22.7$~MeV the data are clearly
   overestimated.

The contribution to a particular kinematically complete breakup conﬁguration,
specified by momenta of three outgoing nucleons, comes from only 
three sets $(p_i, q_i)$ of Jacobi-relative-momenta values,
 belonging to an ellipse of Eq.~(\ref{74}) in  a $q-p$ plane. 
 Performing exclusive breakup measurements one is  restricted only to
points from that curve, what makes the exclusive breakup reaction a
very selective tool.
In contrast to the breakup reaction, averaging over the relative momentum of
 nucleons forming the deuteron causes, that  elastic pd
 scattering gets contributions from a large region in 
 the  $q-p$ plane, which does not overlap with the elliptic curve for 
 the breakup reaction (see Fig.~1 in \cite{wit2020}). 
 It follows that breakup observables should reveal greater sensitivity
 to the underlying dynamics than elastic
pd scattering observables. This motivated us 
 to investigate how the extension of the set
$\left| {\alpha }  \right\rangle$, which has only  a small effect on
elastic scattering observables,   influences the pp QFS cross sections.
  
In Fig.\ref{fig17} we show the pattern of convergence in
partial wave expansion
of the pp QFS cross section around QFS point for some pp QFS configurations
from Fig.\ref{fig16}. 
 The differently colored  dashed lines as well as the solid green line
 are cross sections based on solutions of the Faddeev equation (\ref{63})
  with different partial wave sets and transition amplitude containing all  
  three terms in Eq.~(\ref{79}). The dotted lines show
  the cross sections resulting when only the
  first term in (\ref{79}) was kept and the 
  two Coulomb terms were omitted.
  The same colored lines correspond to the same set of
  $\left| {\alpha }  \right\rangle$-states, with the
  exception of black dotted and
  solid green lines, which correspond to $js3j8$ set. 
  Starting from the set $js3j3$, for which large
  effect of omitting the Coulomb terms is seen and separation
  between blue dotted and
  dashed lines is large, the difference between
  dashed and dotted lines diminishes rapidly with increasing $j$ and
  disappears for $js3j8$ set. Also the
  dashed lines themselves are converging to the prediction obtained
  for the set
  $js3j8$. Including additional partial waves with $j=9$ and higher
  does not change the
  results. It is clear that extending set $\left| {\alpha }  \right\rangle$  
  improves the description of data. That points not only to the need
  for treatment of higher
  partial waves in breakup but the convergence of dotted and dashed results 
  supports also our expectation about the stronger cancellation between
  contributing Coulomb terms with the increasing  number of partial waves.

  Last but not least we  investigate  the
  significance of additional
  two Coulomb contributions to the breakup amplitude of Eq.~(\ref{78})
  (the third and fifth terms) omitted up to now,  
  namely the terms 
$$-\left\langle {\vec p\vec q~} \right|\sum\limits_{\alpha '} {\int
{p'^2 dp'q'^2 dq'\left| {p'q'\alpha '} \right\rangle \left\langle
{p'q'\alpha '} \right|} } t_c^R PG_0   
 \sum\limits_{\alpha
''} {\int {p''^2 dp''q''^2 dq''\left| {p''q''\alpha ''}
\right\rangle \left\langle {p''q''\alpha ''} \right|} } T\left|
{\Phi } \right\rangle $$    and 
$$  \left\langle
{\vec p\vec q~} \right|t_c^R PG_0 \sum\limits_{\alpha '} {\int
{p'^2 dp'q'^2 dq'\left| {p'q'\alpha '} \right\rangle \left\langle
  {p'q'\alpha '} \right|} } T\left| {\Phi } \right\rangle~.$$
 That their contribution can be important is visualised  in
 Figs.~\ref{fig15} and \ref{fig16},  
 where the black double-dotted-dashed lines show the cross section obtained
 when in addition to three contributions in (\ref{79}) also  the term  
$$-\left\langle {\vec p\vec q~} \right|\sum\limits_{\alpha '} {\int
{p'^2 dp'q'^2 dq'\left| {p'q'\alpha '} \right\rangle \left\langle
{p'q'\alpha '} \right|} } t_c^R PG_0   
 \sum\limits_{\alpha
''} {\int {p''^2 dp''q''^2 dq''\left| {p''q''\alpha ''}
\right\rangle \left\langle {p''q''\alpha ''} \right|} } T\left|
            {\Phi } \right\rangle $$ was included.
  The changes of the cross section are quite large 
  at lower energies and become much smaller at $19$ and $22.7$~MeV,
  confirming again diminishing of the Coulomb force effects
  with growing energy. 
 Since changes of the cross section induced by this term are non-negligible, 
it is unavoidable to include in the
transition amplitude  also
 the term
$  \left\langle
{\vec p\vec q~} \right|t_c^R PG_0 \sum\limits_{\alpha '} {\int
{p'^2 dp'q'^2 dq'\left| {p'q'\alpha '} \right\rangle \left\langle
  {p'q'\alpha '} \right|} } T\left| {\Phi } \right\rangle$.
  One would expect that, similarly as for
  the first pair of the two Coulomb terms in (\ref{78}),
  also terms in the second pair would probably
  cancel each other when 
  extending the set $|\alpha>$. To check it requires, 
  however, a calculation of this nontrivial
   modification of the Rutherford term by the strong nucleon-nucleon 
   interactions,  as given
   in Appendix D of Ref. ~\cite{elascoul}
 (Eqs.~(D.6)-(D.8)).
   In Fig.~\ref{fig18} we show the convergence
   pattern in $j$ for the first  pp QFS
   configuration at $13$~MeV, where circles represent
   cross sections obtained with all the 
  five terms included in the breakup amplitude (\ref{78}). To facilitate
  the comparison we also show again by different lines the convergence
  pattern with only three terms in the breakup amplitude (\ref{79}).
  It is seen that indeed circles  converge rapidly to the result
  which  in the maximum of that pp QFS is  
  $\approx 5 \%$ smaller than prediction obtained  with only the first pair
  of Coulomb terms in the breakup transition amplitude. It shows
  that the two contributions
  in  the second  pair of the Coulomb terms of the breakup
  transition amplitude (\ref{78})  do not cancel each other completely
  in the QFS maximum.

  In  Figs.\ref{fig19}, \ref{fig20}, and \ref{fig21}, we show
  converged results
  obtained with $js3j8$ set 
  for all investigated here pp QFS configurations
  at $9.5$ and $13$, $19$ and $22.7$, and $65$~MeV,  respectively.
  At $9.5$ and $13$~MeV the description of data by the
  three-term amplitude  (\ref{79})
  (the green solid line)) is significantly improved when
  compared to $js3j3$ set predictions (the blue short-dashed line).
  Including in addition the second pair of the Coulomb terms
  lowers  by $\approx 5 \%$ 
  cross sections in all the QFS maxima, deteriorating slightly
  the good description of data in that region, leaving it, however, 
  without a change  beyond  the QFS peak region. 
  At the higher energies the contribution from the second pair becomes
  insignificant  and a nice agreement with data at $19$ and $65$~MeV is found.
  The large discrepancies to data at $22.7$~MeV remain.

\section{Summary and conclusions}
\label{sumary}

We formulated and applied a simplified approach to 
 the exact treatment of the pp
 Coulomb force in the momentum space 3N Faddeev calculations, presented
in Refs.~\cite{elascoul} and ~\cite{brcoul}.  
That exact treatment is based on a standard formulation for
 short range forces and
relies on a screening of the long-range Coulomb interaction. 
 It is, however, unhandy for applications since it  contains two 
 contributing terms with a 3-dimensional Coulomb t-matrix,
 which require an unrealistic   amount of computer time and
 resources to compute them in practise. That 
 prevents any application of the exact approach in full-fledged 3N
 calculations.
 Our simplified approach contains the main physical
 ingredients  of the exact method but neglects these complex terms
 altogether,
 relying on their cancellation.
 We have applied it in a wide energy range of the incoming proton, 
 using the AV18 NN potential to  calculate elastic pd scattering
 and breakup observables. The main results are summarized as follows.

\begin{itemize} 

\item[-]
We demonstrated  and showed numerically that the
   elastic pd scattering amplitude has a well defined
   screening limit and therefore does not require any
   renormalization. This is an 
  implication of the fact that only off-shell values of Jacobi momenta and
   consequently, only off-shell 2N t-matrices  
   are required and enter in the determination of that amplitude. 
Well converged elastic pd cross sections and spin observables
 are obtained at finite screening radii. Additional support for the 
claim that infinite R limit is achieved, was provided  
 by using directly the limiting analytical expression
for the 3-dimensional off-shell Coulomb t-matrix when 
calculating the transition amplitude.

\item[-]
Contrary to pd elastic scattering,
for pd breakup it is unavoidable to perform renormalization of the
breakup amplitudes. The reason is that in breakup
only the amplitudes for on-shell
values of Jacobi momenta are required, what demands in consequence
 also half-shell 2N t-matrices. The breakup  
amplitudes have two contributions, one driven by the interaction in the
pp subsystem and second by that in the np subsystem. Only the first part 
requires renormalization. 
 We demonstrated that the renormalization has to be performed
 during  calculation of the breakup transition amplitude and the
 on-shell amplitudes
 themselves can be derived in two different ways which
 lead to the same results.
   We have shown that converged results for 
breakup can be achieved with finite screening radii. The importance
of the renormalization depends on the energy of a 3N system
and on the region in the breakup phase-space.
It is diminishing with the  growing energy, and remains 
 very important at low energies, especially in the region of QFS condition.

\item[-]
  In our approach to breakup two new terms appear. One corresponds
  to the Rutherford pd
  Coulomb amplitude in elastic pd scattering.
  This term was found to be important in the
region of QFS scattering. Calculating that term
with the exact expression for a 3-dimensional half-on-shell  Coulomb t-matrix 
provides an additional test that  the screening
limit for breakup is  achieved.
The second term is a correction to  the first one due to strong
interactions between nucleons. We found that its contribution reduces the 
low energy pp QFS cross sections by $\approx 5 \%$ in the QFS maximum.

\item[-]
  We have checked the validity of the basic assumption underlying our
  simplified approach in the case of a restricted basis of partial
  wave states,
  for which it was possible to compute   the first term with
  a 3-dimensional Coulomb t-matrix in the exact approach.
  Also results for contributions to
  the elastic scattering and breakup transition amplitudes,
  obtained with an extended basis of states,   
  vindicate the cancellation effect between Coulomb terms. 
 The presented results  justify using   our simplified 
 approach, whose requirements for  computer time and
 resources are comparable to standard nd calculations, in future applications.
 This insures that the  pp Coulomb force effects for pd reactions can
 be calculated efficiently and quickly.

\item[-]
We found that large Coulomb force effects are restricted to forward angles
for the elastic pd scattering and to specific regions of the breakup
phase-space. They seem generally to diminish rapidly with the
increasing energy
of the pd system.

\end{itemize}
  
The simplified approach proposed by us can be applied also  in the case
when in addition to pairwise forces a 3NF contributes to the 3N Hamiltonian.  
 Since the structure of Faddeev equations is very similar an extension to
 3N reactions with electromagnetic probes is straightforward.


\clearpage

\acknowledgements
This research was supported in part by the Excellence
Initiative – Research University Program at the Jagiellonian
University in Krak\'ow. One of the authors (J.G.) is grateful to Arnoldas Deltuva for helpful discussions.
The numerical calculations were partly performed on the supercomputers of the JSC, J\"ulich, Germany.

\begin{figure}
\includegraphics[scale=0.55]{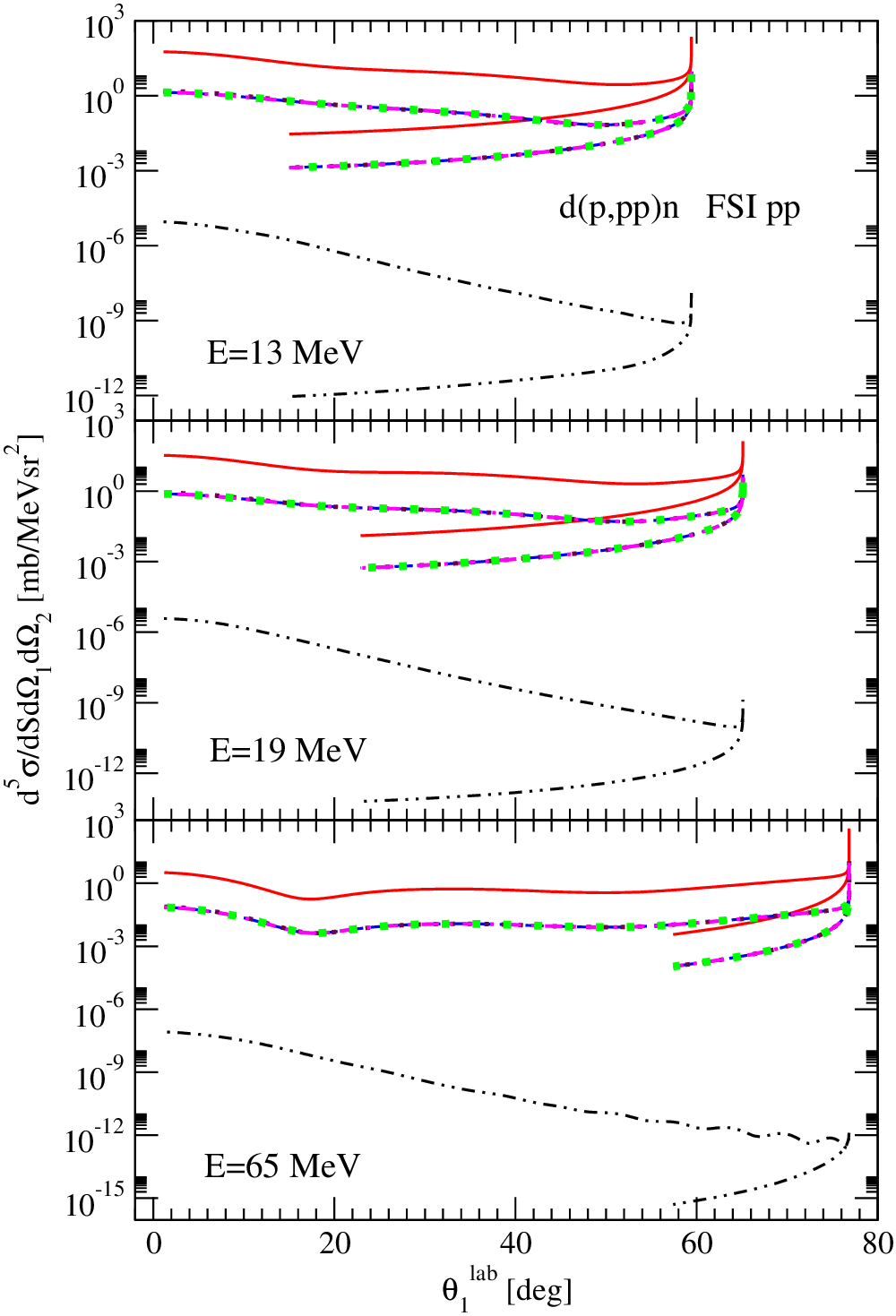}  
\caption{(color online) 
The same as in Fig.~\ref{fig10} but for the  pd  breakup d(p,pp)n 
 cross section $d^5\sigma/d\Omega_1d\Omega_2dS$ calculated  at 
 the pp FSI condition ($\vec p_1=\vec p_2$).
 }
 \label{fig11}
\end{figure}

\begin{figure}
\includegraphics[scale=0.55]{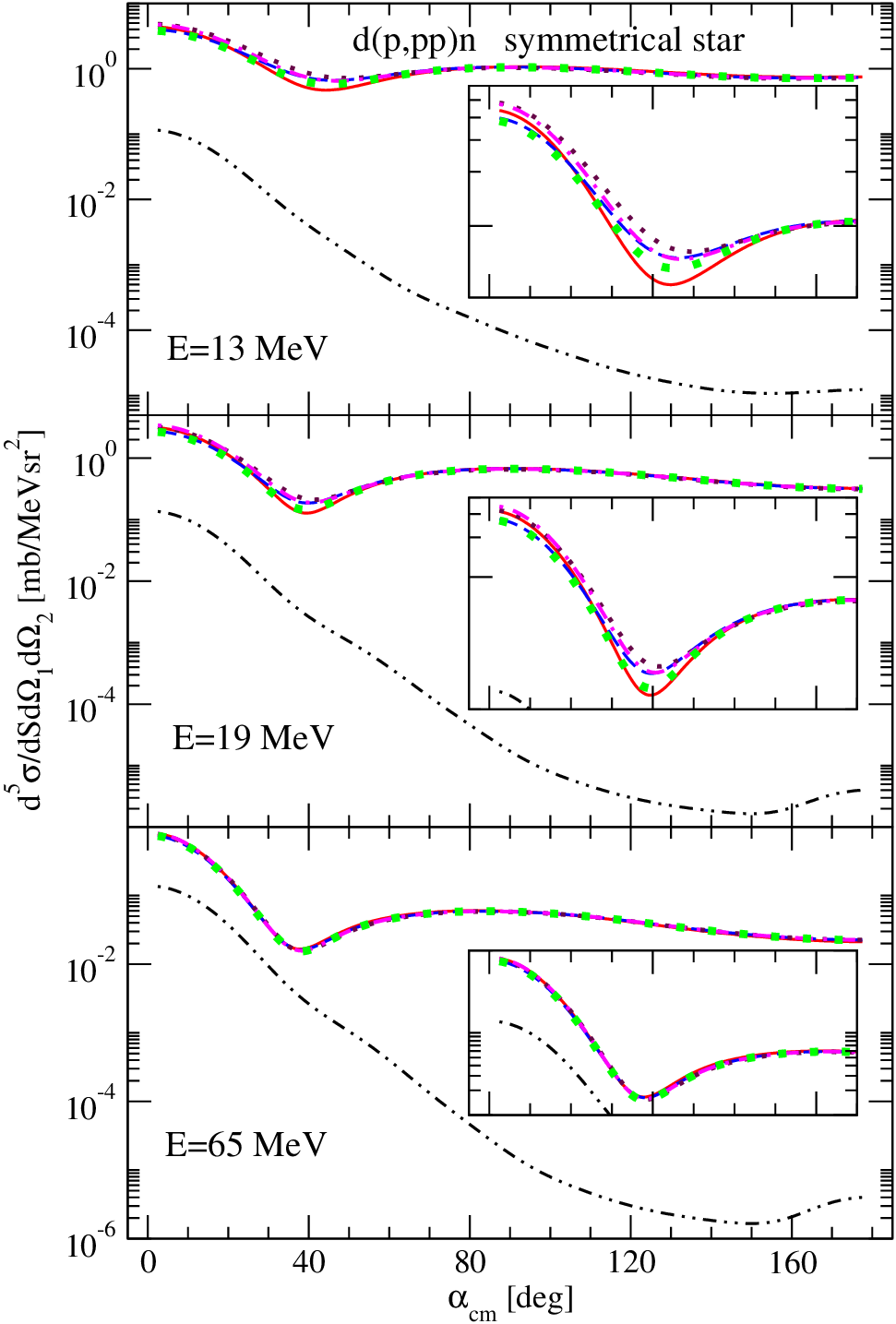}  
\caption{(color online)
The same as in Fig.~\ref{fig10} but for the  pd  breakup d(p,pp)n 
 cross section $d^5\sigma/d\Omega_1d\Omega_2dS$ calculated  at 
 the symmetrical-space-star condition 
 (in the 3N c.m. system the momenta of the three outgoing 
 nucleons are equal and form a symmetric three-point
  star in a plane inclined at an 
 angle  $\alpha_{cm}$ with respect to 
 the incoming proton momentum).
 }
 \label{fig12}
\end{figure}

\begin{figure}
\includegraphics[scale=0.6]{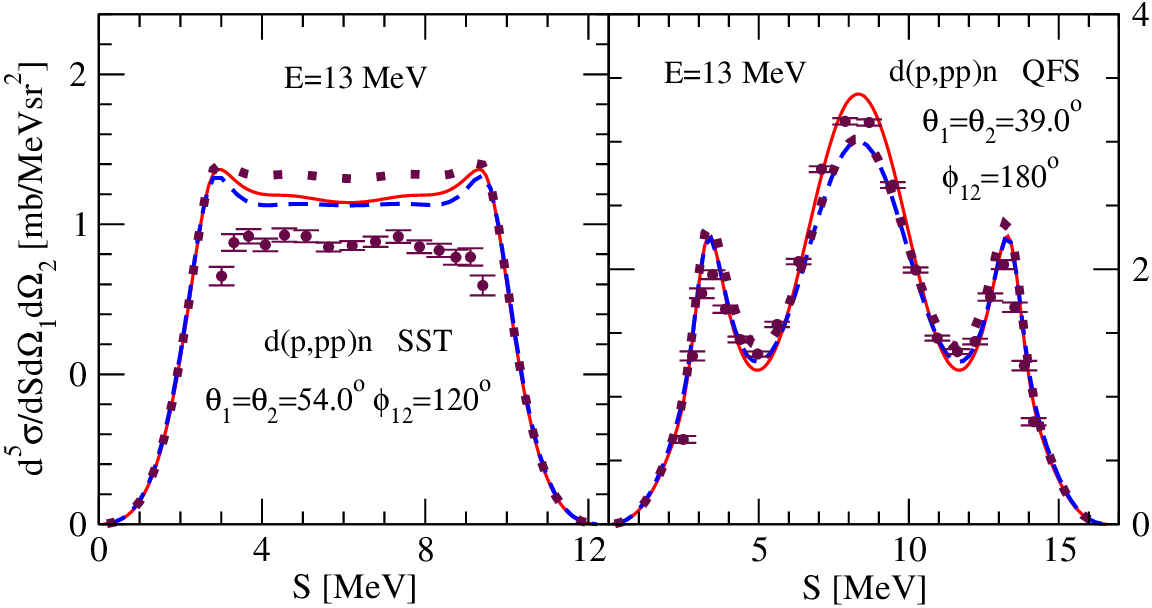}  
\caption{(color online)
The  pd complete breakup d(p,pp)n  
 cross sections $d^5\sigma/d\Omega_1d\Omega_2dS$ 
for the SST and pp QFS configurations at $13$~MeV 
of the incoming proton lab. energy as a function of 
the arc-length of the S-curve.
They are obtained with the screening radius $R=40$~fm ($n=4$)
 and set $js3j3$ partial waves, 
using approach based on Eq.~(\ref{63}) 
(blue short-dashed line) or (\ref{65}) (maroon dotted line). 
 In both approaches the 3-dimensional Coulomb t-matrix 
 is determined according to (\ref{73b}).  
 The on-shell
 Faddeev amplitudes obtained by interpolation from
 the off-shell ones are used 
 and renormalized before calculating observables.
 The red solid line is the corresponding nd breakup cross section.
 The maroon circles are pd data from Ref.~\cite{exp2}. 
 }
 \label{fig13}
\end{figure}

\begin{figure}
 \includegraphics[scale=0.65]{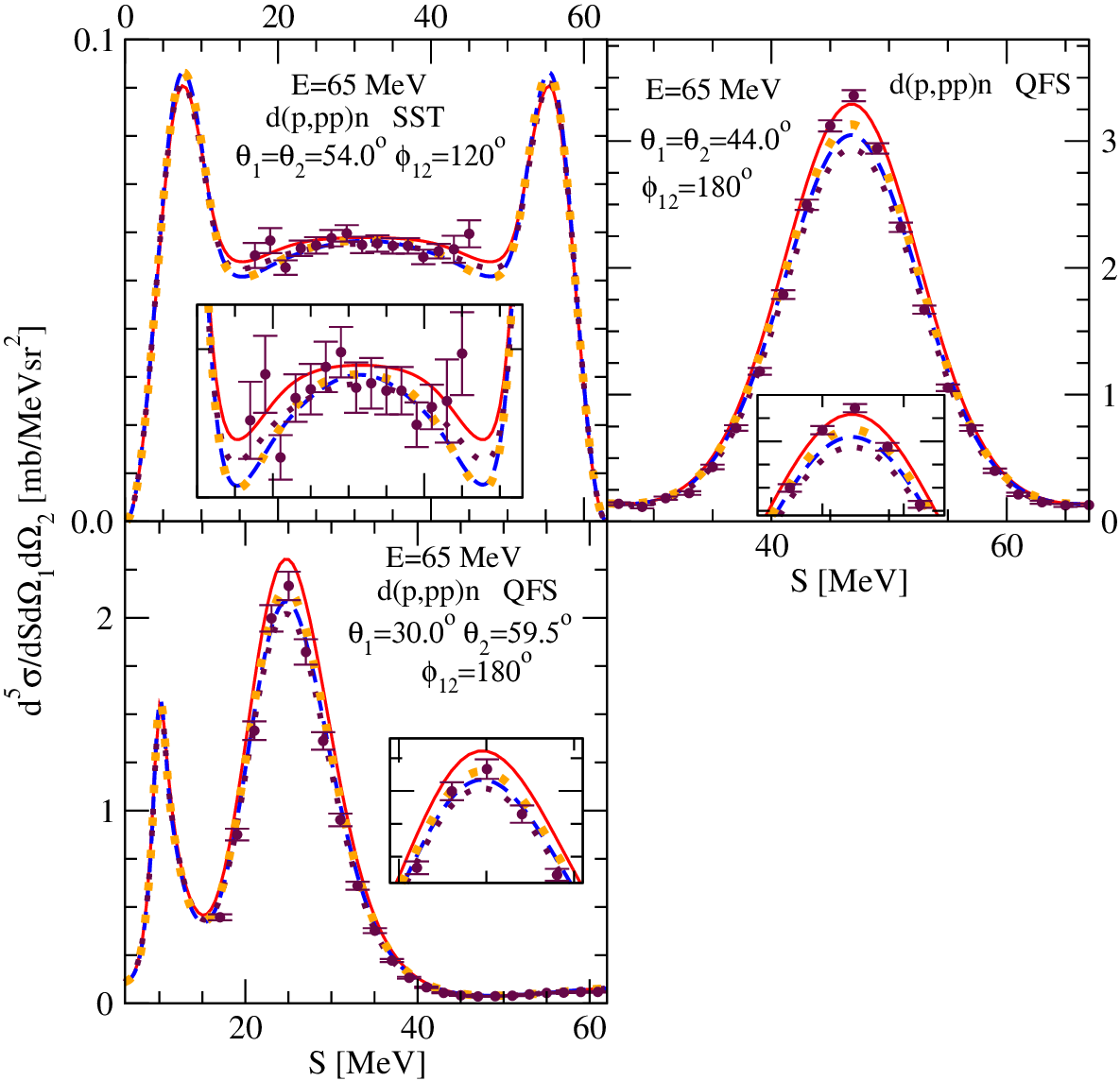}  
\caption{
The same as in Fig.~\ref{fig13} but for the  pd complete breakup d(p,pp)n 
cross sections $d^5\sigma/d\Omega_1d\Omega_2dS$  in SST and pp QFS at $65$~MeV.
Here in both approaches the 3-dimensional Coulomb t-matrix 
was determined by solving the 3-dimensional Lippmann-Schwinger
equation with the screening radius $R=40$~fm ($n=4$). In case of the
approach based on Eq.~(\ref{63}) effects of using
3-dimensional Coulomb t-matrix determined according to (\ref{73b}) is shown
by the orange dotted lines. 
The maroon circles are pd data from Ref.~\cite{exp5}.
 }
 \label{fig14}
\end{figure}

\begin{figure}
\includegraphics[scale=0.6]{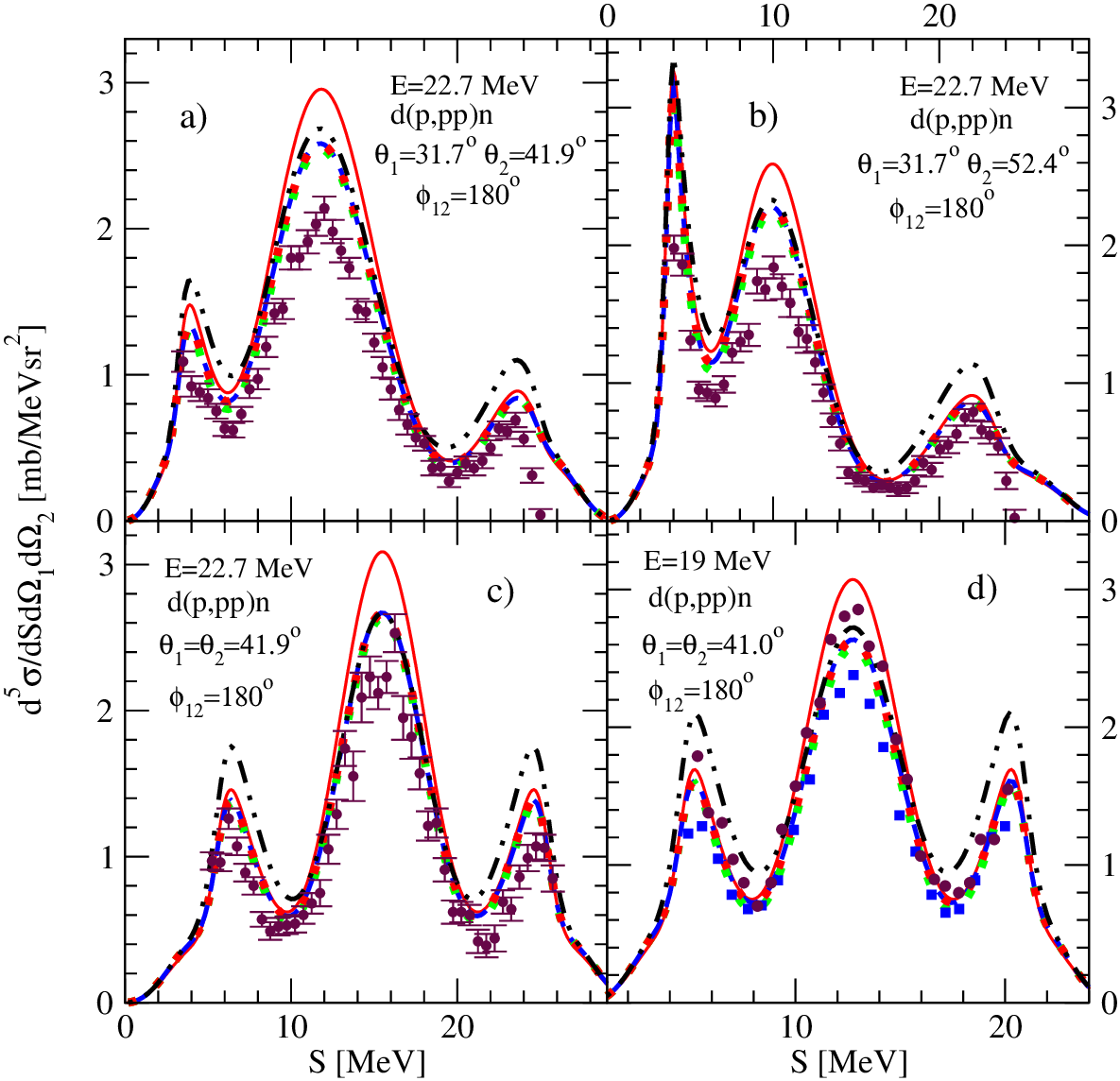}  
\caption{(color online) 
The  pd complete breakup d(p,pp)n  
 cross sections $d^5\sigma/d\Omega_1d\Omega_2dS$ 
 for pp QFS configurations at $22.7$~MeV ( a), b), and c) )
 and $19$~MeV ( d) )
of the incoming proton lab. energy, as a function of 
the arc-length of the S-curve.
 They were obtained with the screening radius $R=40$~fm ($n=4$) 
using approach based on Eq.~(\ref{63}) 
 and set $js3j3$ of partial waves.  
 The on-shell
 Faddeev amplitudes obtained by interpolation from the off-shell ones  
  were used 
 and renormalized before calculating observables with amplitude
 of Eq.~(\ref{79}) (blue short-dashed line).
 The green dotted line is a result  without renormalization.
 The black 
  double-dotted-dashed line shows the result when also the third term
  in (\ref{78}) is included in the breakup transition amplitude.
 The red  dotted line  is obtained  with
 the limiting $t_c$ of Eq.~(\ref{73b}). 
 The red solid line is the corresponding nd breakup cross section.
 The maroon circles are pd data from Ref.~\cite{miljanic}
 for a), b), and c),
 and from Ref.~\cite{exp3} for d). The blue squares in d) are
 pd data from Ref.~\cite{exp3}. 
 }
 \label{fig15}
\end{figure}

\begin{figure}
\includegraphics[scale=0.58]{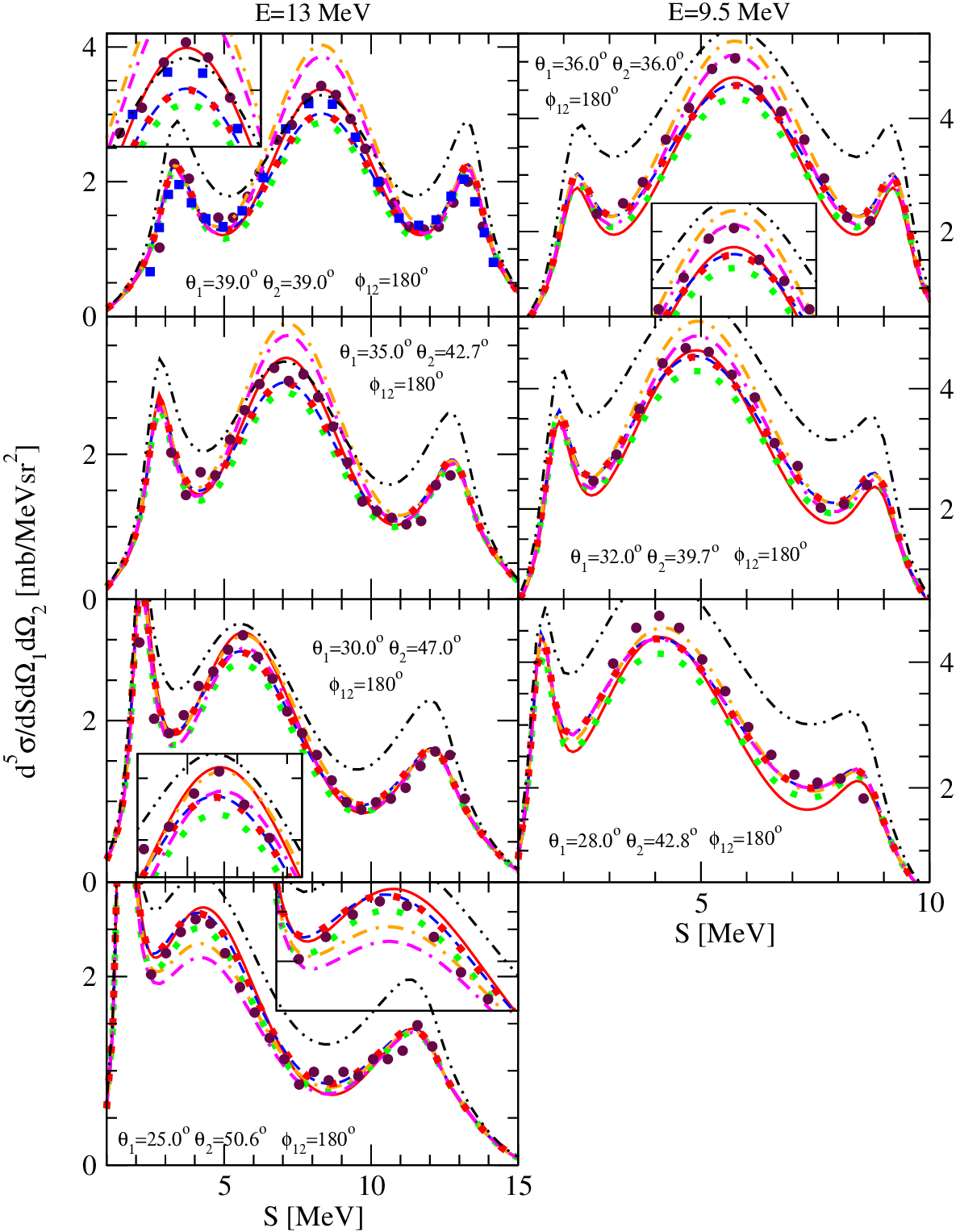}  
\caption{(color online) 
  The  same as in Fig.~\ref{fig15} but for the
 pp QFS configurations at $13$~MeV and $9.5$~MeV.
 The orange dashed-dotted and 
  magenta double-dashed-dotted lines are renormalized and unrenormalized
  results, respectively, when  both Coulomb terms are omitted in the breakup
  transition amplitude (\ref{79}).
  The red solid line is the nd breakup cross section.  
 The maroon circles are pd data from Ref.~\cite{sagara_qfs} and blue squares
 at $13$~MeV are pd data from Ref.~\cite{schieck}. 
 }
 \label{fig16}
\end{figure}

\begin{figure}
\includegraphics[scale=0.7]{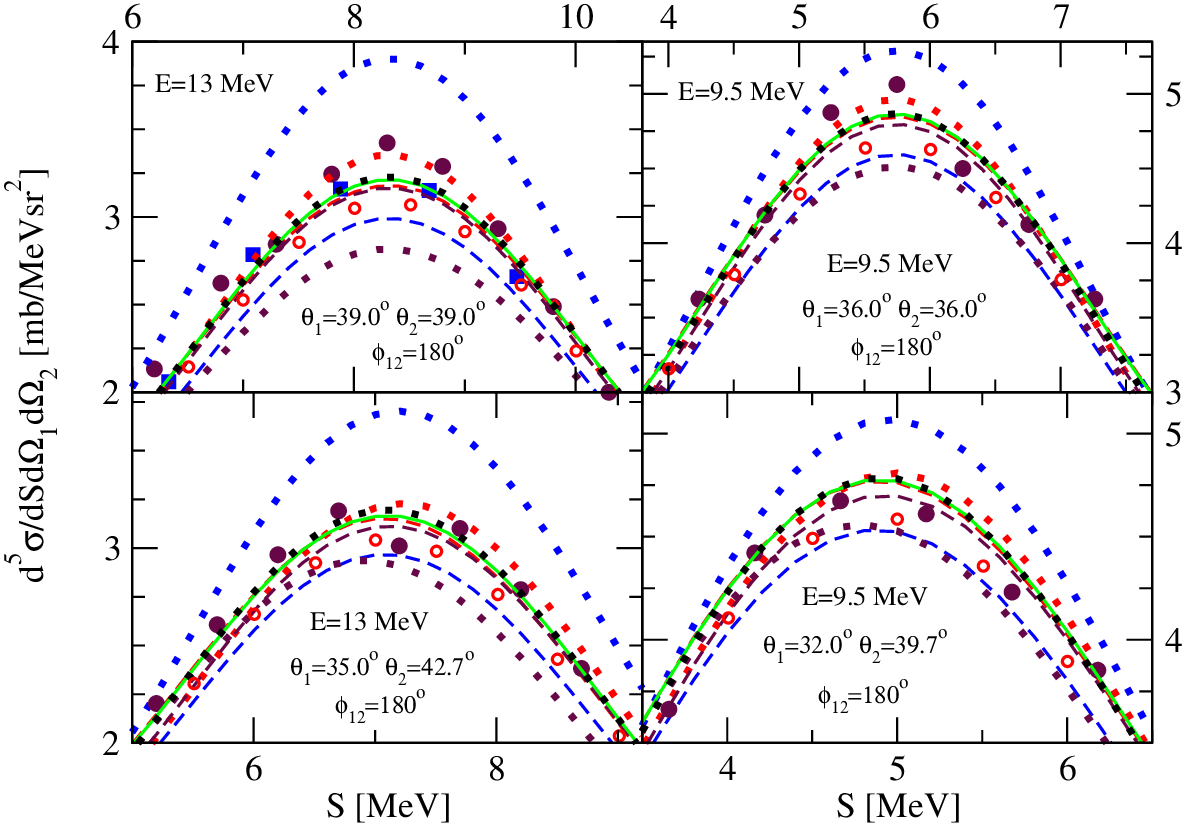}  
\caption{(color online) 
  The pattern of convergence in the sets of
  partial waves around the QFS pp condition
  for cross sections in chosen configurations from Fig.\ref{fig16},
  calculated with  the amplitude of Eq.~(\ref{79}).
 The meaning of the  data sym\-bol\-s 
 is the same as in Fig.~\ref{fig16}.
 Equally  colored dashed and
 dotted lines correspond to identical set of partial waves. For dashed lines
 all  three terms in Eq.~(\ref{79}) 
 contribute to the breakup amplitude while for dotted lines
 only the first term in (\ref{79}) was taken into account.
 The sets of partial waves are:
 blue - $js3j3$, maroon - $js3j5$, red - $js3j7$.
 For the set $js3j8$ the color of
 lines are solid green  and black dotted for the case when all
 three and only the first term were taken into account, respectively.  
 The red circles show the cross sections   calculated with the set js3j8  
	and all five terms of Eq.~(\ref{78}) included in the breakup amplitude.
}
 \label{fig17}
\end{figure}

\begin{figure}
\includegraphics[scale=1.0]{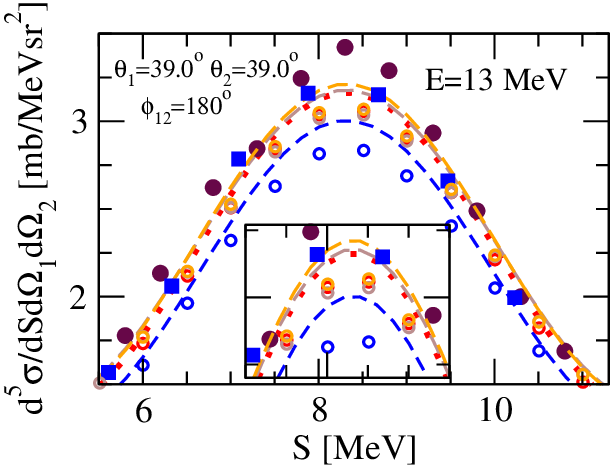}  
\caption{(color online) 
  The pattern of convergence in the sets of
  partial waves around the QFS pp condition
  for cross section of the first  configuration at $E=13$~MeV
  from Fig.\ref{fig16},
  calculated with  the amplitude of Eq.~(\ref{79}) (lines)
  and (\ref{78}) (circles). 
 The meaning of the  data sym\-bol\-s 
 is the same as in Fig.~\ref{fig16}.
 Equally  colored lines and circles 
 correspond to identical set of partial waves. For  lines
 all  three terms in (\ref{79}) and for circles all five terms
 in (\ref{78})
 contribute to the breakup amplitude. 
  The sets of partial waves are:
  blue (short-dashed) - $js3j3$, red (dotted) - $js3j5$,
  brown (long dashed) - $js3j7$, and orange (long dashed) - $js3j8$.
}
 \label{fig18}
\end{figure}

\begin{figure}
\includegraphics[scale=0.56]{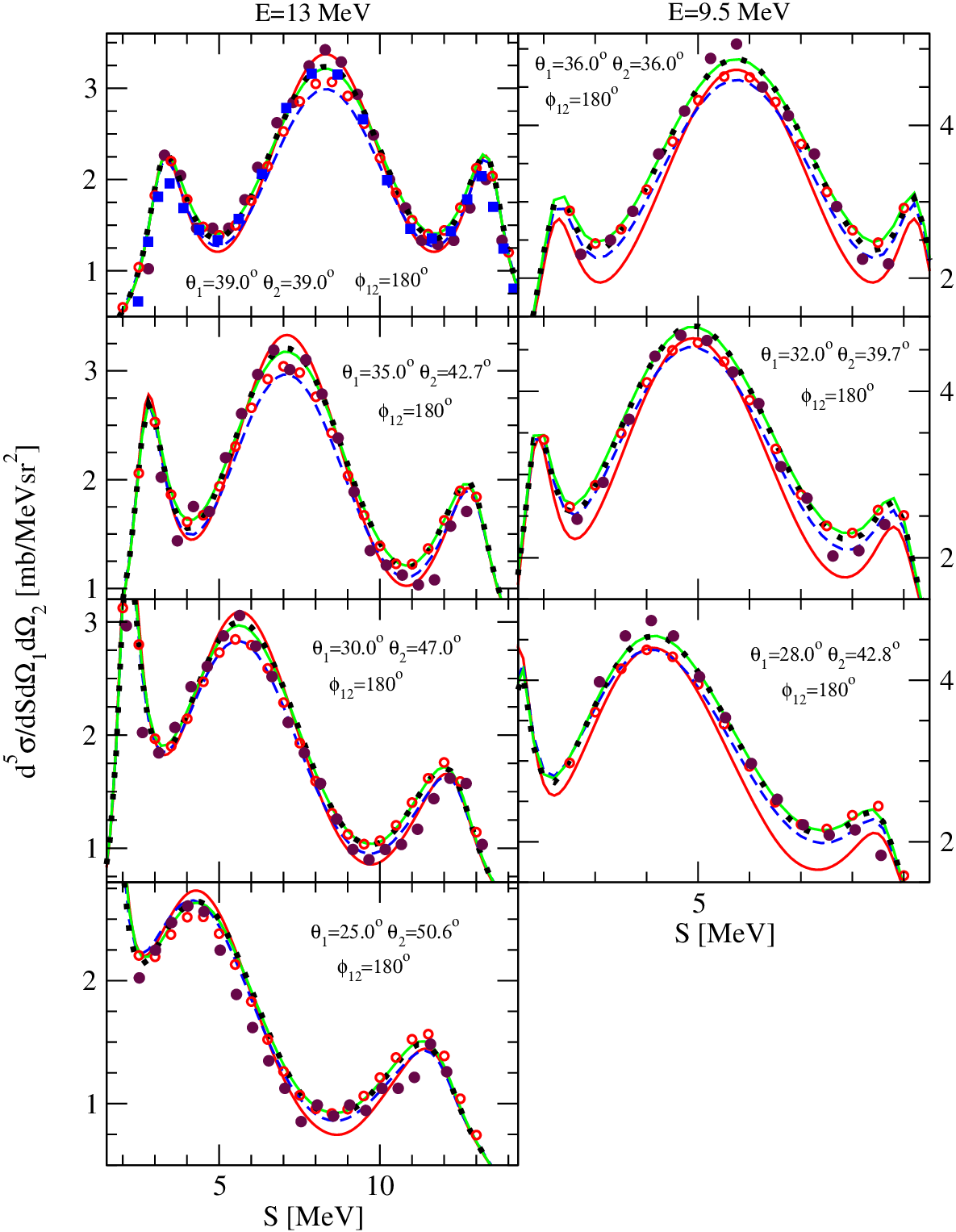}  
\caption{(color online) 
The  pd complete breakup d(p,pp)n  
 cross sections $d^5\sigma/d\Omega_1d\Omega_2dS$ 
 for the configurations from Fig.~\ref{fig16},  calculated 
 using approach based on Eq.~(\ref{63}) 
 and set $js3j8$ of partial waves,
 with on-shell
 Faddeev amplitudes obtained by interpolation from the off-shell ones,
 renormalized before calculating observables.
 The screening radius $R=40$~fm ($n=4$) was used and the green solid lines 
 show  the results with all three terms in (\ref{79})
 contributing to the breakup amplitude, while for the black 
 dotted lines  only the first term was taken into account.  
 The red circles show the cross sections   calculated with all five terms
  in (\ref{78})  included in the breakup amplitude.  
  For the sake of comparison also results with the set $js3j3$
  and three terms in (\ref{79}) contributing to the breakup amplitude,  
 are shown (the blue short-dashed line).
The red solid line is the corresponding nd cross section.
 The meaning of the  data sym\-bol\-s 
 is the same as in Fig.~\ref{fig16}. 
}
 \label{fig19}
\end{figure}

\begin{figure}
\includegraphics[scale=0.7]{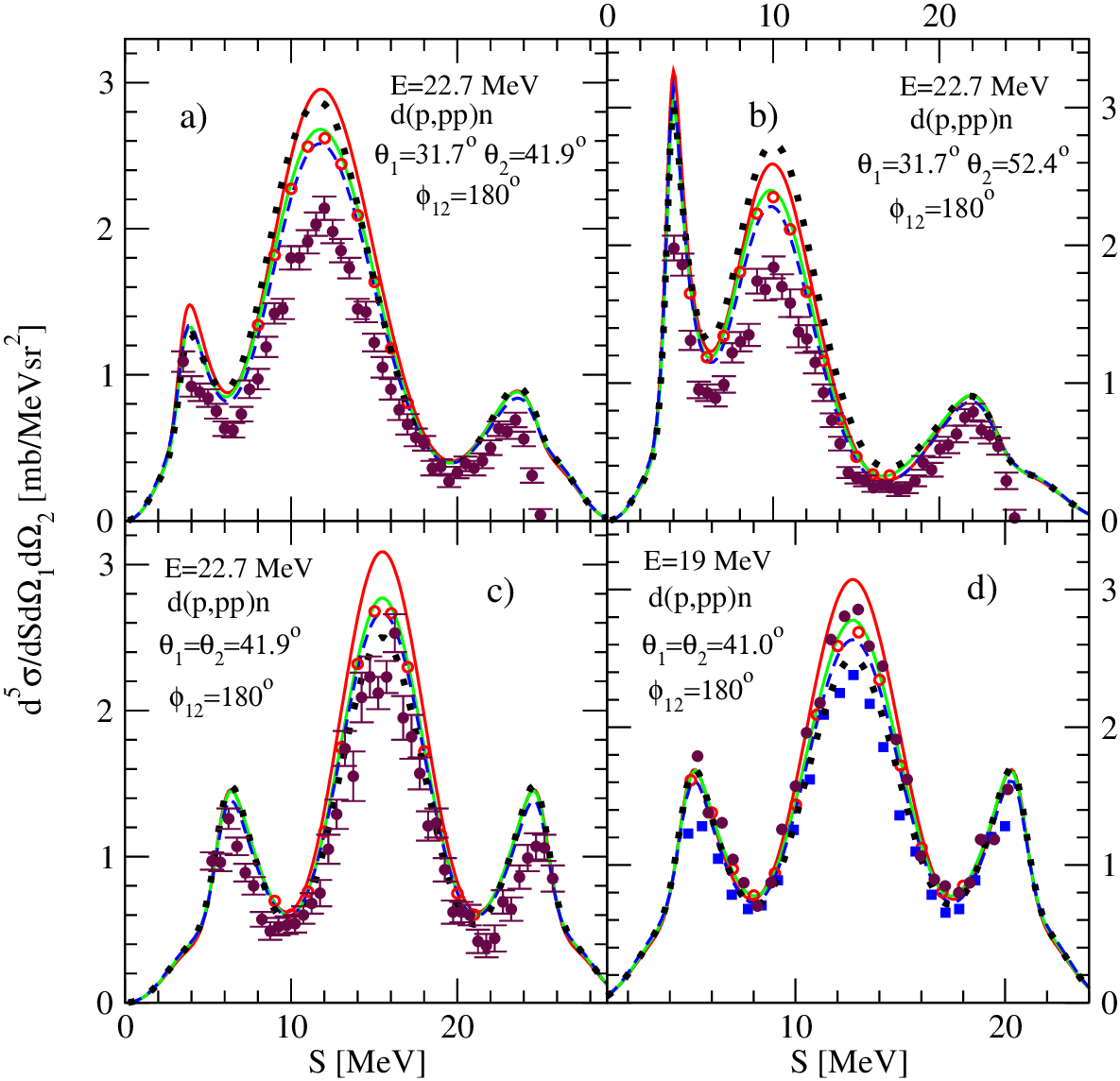}  
\caption{(color online)
  The same as in Fig.~\ref{fig19} but for the
  configurations from Fig.~\ref{fig15}. 
}
 \label{fig20}
\end{figure}

\begin{figure}
\includegraphics[scale=0.7]{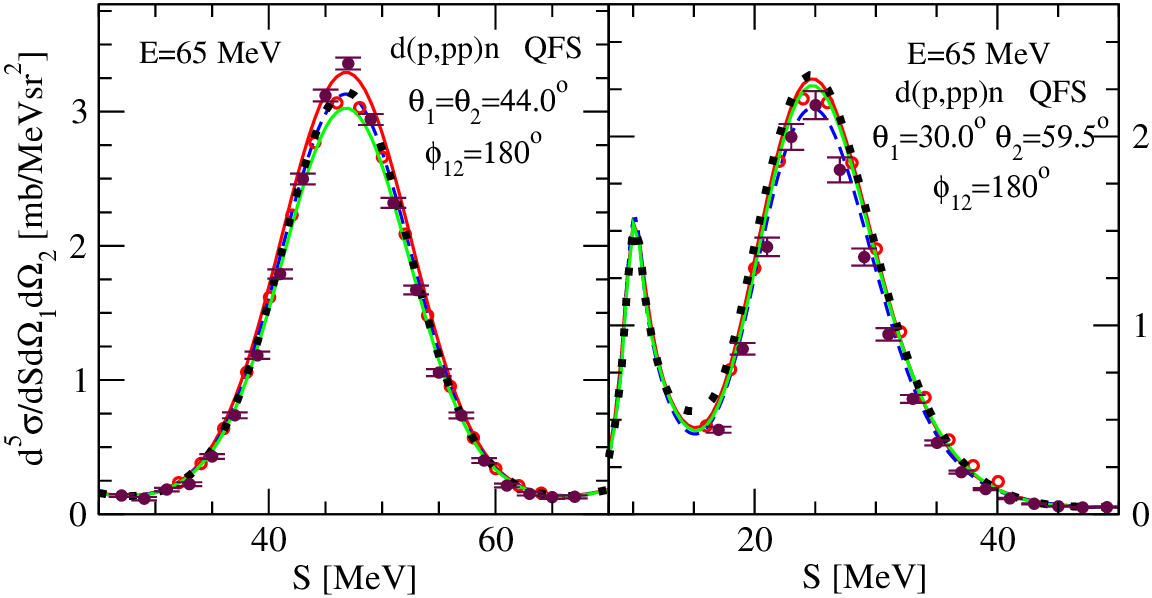}  
\caption{(color online) 
  The same as in Fig.~\ref{fig19} but for the pp QFS configurations
  from Fig.~\ref{fig14}.
}
 \label{fig21}
\end{figure}


\begin{thebibliography}{99}

\bibitem{Alt78} E. O. Alt, W. Sandhas, and H. Ziegelmann, Phys. Rev. {\bf C
17}, 1981 (1978).

\bibitem{Alt96} E. O. Alt and W. Sandhas, in Coulomb Interactions in Nuclear
and Atomic Few-Body Collisions, ed. by F.S. Levin and D. Micha (Plenum,
New York 1996), p.1.

\bibitem{Alt94} E. O. Alt and M. Rauh,  Phys. Rev. {\bf C49}, R2285 (1994).


\bibitem{Alt2002} E. O. Alt, A. M. Mukhamedzhanov, M. M. Nishonov, and
A. I. Sattarov, Phys. Rev. {\bf C65}, 064613 (2002).

\bibitem{physrep96}  W. Gl\"ockle, H. Wita{\l}a, D. H\"uber, H. Kamada, J.
Golak,
 Phys. Rep. {\bf{274}}, 107 (1996).


\bibitem{kievski} A. Kievsky, M. Viviani, and S. Rosati, Phys. Rev. {\bf C
52}, R15 (1995).

\bibitem{delt2005br} A. Deltuva, A. C. Fonseca, and P. U. Sauer,
  Phys. Rev.  C{\bf{72}}, 054004 (2005).

\bibitem{delt2005el} A. Deltuva, A. C. Fonseca, and P. U. Sauer,
  Phys. Rev. C{\bf{71}}, 054005 (2005).

\bibitem{delt2006br} A. Deltuva, A. C. Fonseca, and P. U. Sauer,
  Phys. Rev.  C{\bf{73}}, 057001 (2006).   
  
\bibitem{stephan} E. Stephan et al.,
 Phys. Rev. C{\bf{76}}, 057001 (2007).

\bibitem{elascoul} H. Wita{\l}a, R. Skibi\'nski, J. Golak,
  W. Gl\"ockle,
  Eur. Phys. Journal A{\bf{41}}, 369 (2009).

\bibitem{brcoul} H. Wita{\l}a, R. Skibi\'nski, J. Golak,
  W. Gl\"ockle,
  Eur. Phys. Journal A{\bf{41}}, 385 (2009).  

\bibitem{wit91} H. Wita{\l}a, W. Gl\"ockle, H.Kamada,
  Phys. Rev. C{\bf{43}}, 1619  (1991).

\bibitem{chen72} J.C.Y. Chen and A.C. Chen,
in Advances of Atomic and Molecular Physics,
 edited by D. R. Bates and J. Estermann ( Academic, New York, 1972), Vol. 8.

\bibitem{kok1980} L. P. Kok, H. van Haeringen,  P
  hys. Rev. C{\bf{21}}, 512 (1980).  
  
\bibitem{gloeckle83} W. Gl\"ockle, The Quantum Mechanical Few-Body Problem,
  Springer Verlag 1983.

\bibitem{ford1964} W.F. Ford, Phys. Rev. {\bf 133}, B1616 (1964).


\bibitem{ford1966} W.F. Ford, J. Math. Phys. {\bf 7}, 626 (1966).

\bibitem{kamada05} M. Yamaguchi, H. Kamada, and Y. Koike, Prog. Theor. Phys.
{\bf 114} , 1323 (2005).

\bibitem{taylor1} J.R. Taylor,  Nuovo Cimento {\bf B23}, 313 (1974).

\bibitem{taylor2} M.D. Semon and J.R. Taylor,
  Nuovo Cimento {\bf A26}, 48 (1975).

\bibitem{abr_steg}  M. Abramowitz, I.A. Stegun, (Editors) Handbook of
  Mathematical Functions,  (Dover Publ., N.Y., 1972).
  
\bibitem{skib2009}  R. Skibi\'nski, J. Golak, H. Wita{\l}a,  and W.Gl\"ockle, 
Eur. Phys. Journal A{\bf{40}}, 215  (2009).

\bibitem{kok1981} L. P. Kok, H. van Haeringen,
  Phys. Rev. Lett. {\bf{46}}, 1257 (1981).   

\bibitem{av18}  R. B. Wiringa, V. G. J. Stoks, R. Schiavilla, 
  Phys. Rev. C{\bf 51}, 38 (1995).

\bibitem{epel2020}  E. Epelbaum et al., 
  Eur. Phys. J. {\bf A 56}, 92 (2020), and references therein.

\bibitem{wit2022}  H. Wita{\l}a et al., 
  Phys. Rev. C{\bf 105}, 054004 (2022), and references therein.

\bibitem{wit2020}  H. Wita{\l}a, J.Golak, R.Skibi\'nski,
  V. Soloviov, K.Topolnicki,
and V. Urbanevych, Phys. Rev. C{\bf 101}, 054002 (2020).

\bibitem{haer1985} H. van Haeringen, Charged Particle Interactions,
  Theory and Formulas, (Coulomb Press, Leyden, 1985).
    

\bibitem{grub12} W.~Gr\"uebler et al.,
  Nucl. Phys. A{\bf 398}, 445 (1983).

\bibitem{schieck} G. Rauprich et. al., Few-Body Syst. {\bf{5}}, 67 (1988).

\bibitem{sagara} K.~Sagara {\it et al.},
  Phys. Rev. C {\bf 50}, 576 (1994).

\bibitem{shimizu} H. Shimizu {\em et al}.,  Nucl. Phys. A {\bf 382}, 
  242 (1982).

\bibitem{sak99}  H. Sakai  {\em et al}.,   
Phys. Rev. Lett. {\bf 84}, 5288 (2000).

\bibitem{hatanaka} K. Hatanaka {\em et al}., Phys. Rev. C {\bf 66}, 
044002 (2002).

\bibitem{maeda250} Y.~Maeda {\it et al.},
  Phys. Rev. C {\bf 76}, 014004 (2007).
  
\bibitem{sperisen84} F.~Sperisen {\it et al.},
  Nucl. Phys. A {\bf 422}, 81 (1984).

\bibitem{cameron}  H. Wita{\l}a et al., Few-Body Syst. {\bf{15}}, 67 (1993).

\bibitem{seki140} K. Sekiguchi {\em et al}., Phys. Rev. C {\bf 65}, 
034003 (2002).

\bibitem{seki250} 
K.\ Sekiguchi {\it et al.}, Phys.\ Rev.\ C {\bf 89}, 064007 (2014).

\bibitem{sawada83} M.~Sawada {\it et al.},
  Phys. Rev. C {\bf 27}, 1932 (1983).

\bibitem{ermisch} K.~Ermisch {\it et al.},
  Phys. Rev. C {\bf 71}, 064004 (2005).
  
\bibitem{exp2} G. Rauprich et al., Nucl. Phys. {\bf{A535}}, 313 (1991).

\bibitem{exp5} M. Allet  et al.,  Few-Body Syst. C{\bf{20}}, 27 (1996).

\bibitem{miljanic}  M. Zadro et al., Il Nuovo Cimento
  {\bf{107A}}, 185 (1994).

\bibitem{exp3} H. Patberg et al.,  Phys. Rev. C{\bf{53}}, 1497 (1996).

\bibitem{sagara_qfs}  Y. Eguchi et al.,  EPJ Web
  of Conferences 3, 04007 (2010),
    DOI:10.1051/epjconf/20100304007.


\end{thebibliography}
\end{document}